\documentclass[]{iopart}
\usepackage{graphicx, epsfig} %include figure files
\usepackage{bm}

\expandafter\let\csname equation*\endcsname\relax
\expandafter\let\csname endequation*\endcsname\relax

\usepackage[caption=false]{subfig}
\usepackage[usenames]{xcolor}

\usepackage{marvosym}
\usepackage{yfonts}
\usepackage{upgreek}
\usepackage{pdflscape}
\definecolor{linkcolor}{rgb}{0.0, 0.3, 0.5}
\definecolor{purple}{rgb}{0.7, 0.05, 0.5}
\usepackage[unicode,colorlinks=true,citecolor=linkcolor,linkcolor=linkcolor,urlcolor=linkcolor]{hyperref}
\usepackage[all]{hypcap}
\usepackage{amsmath,amsfonts,amssymb}
\usepackage{bm}
\usepackage{longtable}
\usepackage[T1]{fontenc}
\usepackage[utf8]{inputenc}
\usepackage{verbatim}
\usepackage{pifont}
\usepackage{yfonts}
\usepackage{url}
\usepackage{multirow}
\usepackage{tensor}
\usepackage{mathtools}
\usepackage[normalem]{ulem}
\usepackage{fancyhdr}
\usepackage{xcolor}
\usepackage{soul}

\renewcommand{\vec}[1]{\boldsymbol{#1}}
\newcommand{\iu}{\mathrm{i}\mkern1mu}
\newcommand{\du}{\mathrm{d}}

\newcommand*{\defeq}{\mathrel{\vcenter{\baselineskip0.5ex \lineskiplimit0pt
                     \hbox{\scriptsize.}\hbox{\scriptsize.}}}%
                     =}
\newcommand*{\invdefeq}{=\mathrel{\vcenter{\baselineskip0.5ex \lineskiplimit0pt
                     \hbox{\scriptsize.}\hbox{\scriptsize.}}}%
                     }

\begin{document}
\newcounter{count}

\pagestyle{fancy}\lhead{Malaise and remedy of binary boson-star initial data}
\chead{}
\rhead{\thepage}
\lfoot{}
\cfoot{}
\rfoot{}

\begin{center}
\title{\large Malaise and remedy of binary boson-star initial data}
\end{center}

\author{
Thomas Helfer$^{1}$,
Ulrich Sperhake$^{1,2,3}$,
Robin Croft$^{2}$,
Miren Radia$^{2}$,
Bo-Xuan Ge$^4$,
Eugene A. Lim$^4$
}

\address{$^{1}$~Department of Physics and Astronomy, Johns Hopkins University, 3400 N. Charles Street, Baltimore, Maryland 21218, USA}

\address{$^{2}$~Department of Applied Mathematics and Theoretical Physics,
Centre for Mathematical Sciences, University of Cambridge,
Wilberforce Road, Cambridge CB3 0WA, United Kingdom}

\address{$^{3}$~Theoretical Astrophysics 350-17,
California Institute of Technology,
1200 E California Boulevard, Pasadena, CA 91125, USA}

\address{$^{4}$~Theoretical Particle Physics and Cosmology Group, Physics Department,Kings College London, Strand, London WC2R 2LS, United Kingdom}

\ead{thomashelfer@live.de,
     U.Sperhake@damtp.cam.ac.uk,
     M.R.Radia@damtp.cam.ac.uk,
     bo-xuan.ge@kcl.ac.uk,
     eugene.a.lim@gmail.com
    }

\begin{abstract}
Through numerical simulations of boson-star head-on collisions, we
explore the quality of binary initial data obtained from the
superposition of single-star spacetimes.  Our results demonstrate
that evolutions starting from a plain superposition of individual
boosted boson-star spacetimes are vulnerable to significant unphysical
artefacts. These difficulties can be overcome with a simple
modification of the initial data suggested in \cite{Helfer:2018vtq}
for collisions of oscillatons.
While we specifically consider massive complex scalar field boson
star models up to a 6th-order-polynomial potential, we argue that
this vulnerability is universal and present in other kinds of exotic
compact systems and hence needs to be addressed.
\end{abstract}

\maketitle

%=============================================================================
\section{Introduction}
\label{sec:intro}

The rise of gravitational-wave (GW) physics as an observational
field, marked by the detection of GW150914 \cite{Abbott:2016blz}
and followed by about 50 further compact binary events
\cite{LIGOScientific:2018mvr,LIGOScientific:2020ibl} over the past
years, has opened up unprecedented opportunities to explore
gravitational phenomena. From tests of general relativity
\cite{Berti:2015itd,TheLIGOScientific:2016src,Abbott:2018lct,LIGOScientific:2019fpa,LIGOScientific:2020tif,Moore:2021eok}
to the exploration of BH populations
\cite{Trifiro:2015zda,Belczynski:2017gds,LIGOScientific:2020kqk,Baibhav:2020xdf,Gerosa:2021mno}
or charting the universe with independent new methods
\cite{LIGOScientific:2017adf,LIGOScientific:2019zcs}, GW astronomy
offers potential for revolutionary insight into long-standing open
questions; for a review see \cite{Barack:2018yly}. Some answers,
such as the association of a soft gamma-ray burst with the neutron
star merger GW170817 \cite{TheLIGOScientific:2017qsa,Monitor:2017mdv}
have already raised our understanding to new levels. GW physics
furthermore establishes new concrete links to other fields of
research, most notably to particle and high-energy physics and the
exploration of the dark sector of the universe
\cite{Cardoso:2014uka,Barack:2018yly}. Two important ingredients
of this remarkable connection are the characteristic interaction
of fundamental fields with compact objects through superradiance
\cite{Brito:2015oca} and their capacity to form compact objects
through an elaborate balance between the intrinsically dispersive
character of the fields and their self-gravitation. The latter
feature has given rise to the hypothesis of a distinct class of
compact objects as early as the 1950s \cite{Wheeler:1955zz}. In
contrast to their well known fermionic counterparts -- stars, white
dwarfs or neutron stars -- these compact objects are composed of
bosonic particles or fields and, hence, commonly referred to as
{\em Boson Stars} (BS). GW observations provide the first systematic
approach to search for populations of these objects or to constrain
their abundance. As with all other GW explorations, the success of
this exploration is heavily reliant on the availability of accurate
theoretical predictions for the anticipated GW signals.  This type
of calculation, using numerical relativity techniques
\cite{Baumgarte:2021skc}, is the topic of this work.

The idea of bosonic stars dates back to Wheeler's 1955 study of
gravitational-electromagnetic entities or {\it geons}
\cite{Wheeler:1955zz}.  By generalising from real to complex-valued
fundamental fields, it is even possible to obtain genuinely stationary
solutions to the Einstein-matter equations. First established for
spin 0 or scalar fields \cite{Feinblum:1968nwc,Kaup:1968zz,Ruffini:1969qy},
this idea has more recently been extended to spin 1 or vector (aka
{\it Proca}\footnote{Even though the term ``boson star''
generally applies to compact objects formed of any bosonic
fields, it is often used to specifically denote stars made
up of a {\it scalar} field. Stars
composed of vector fields, in contrast, are most commonly
referred to as {\it Proca} stars. Unless specified otherwise,
we shall accordingly assume the term boson star to imply
scalar-field matter.})
fields \cite{Brito:2015pxa} as well as wider classes of scalar BSs
\cite{Alcubierre:2018ahf,Choptuik:2019zji}.  In the wake of the
dramatic progress of numerical relativity in the simulations of
black holes (BHs) \cite{Pretorius:2005gq,Campanelli:2005dd,Baker:2005vv}
(see \cite{Sperhake:2014wpa} for a review), the modelling of BSs
and binary systems involving BSs has rapidly gathered pace.

The first BS models computed in the 1960s consisted of a massive
but non-interacting complex scalar field $\varphi$.  This class of
stationary BSs, commonly referred to as {\it mini boson stars},
consists of a one parameter family of ground-state solutions
characterised by the central scalar-field amplitude that reveals a
stability structure analogous to that of Tolman-Oppenheimer-Volkoff
\cite{Tolman:1939jz,Oppenheimer:1939ne} stars: a stable and an
unstable branch of ground-state solutions are separated by the
configuration with maximal mass
\cite{Breit:1983nr,Gleiser:1988ih,Seidel:1990jh}. For each ground-state
model, there furthermore exists a countable hierarchy of excited
states with $n>0$ nodes in the scalar profile
\cite{Lee:1991ax,Jetzer:1991jr,Liddle:1992fmk}. Numerical evolutions
of these excited BSs demonstrate their unstable character, but also
reveal significant variation in the instability time scales
\cite{Balakrishna:1997ej}.

Whereas mini BS models are limited in terms of their maximum
compactness, self-interacting scalar fields can result in significantly
more compact stars, even denser than neutron stars
\cite{Colpi:1986ye,Lee:1986ts,Schunck:1999zu,Hartmann:2012da}. This
raises the intriguing question whether compact BS binaries may
reveal themselves through characteristic GW emission analogous to
that from BHs or NSs \cite{Bustillo:2020syj}. Recent studies conclude
that this may well be within the grasp of next-generation GW detectors
and, in the case of favourable events, even with advanced LIGO
\cite{Sennett:2017etc,DiGiovanni:2020ror,Toubiana:2020lzd}.

One of the characteristic properties of BSs is the quantised nature
of their spin. The linearised Einstein equations in the slow-rotation
limit lead to a two-dimensional Poisson equation that does not admit
everywhere regular solutions except for trivial constants; in
consequence BSs cannot rotate perturbatively \cite{Kobayashi:1994qi}.
By relaxing the slow-rotation approximation, Schunck and Mielke
\cite{Schunck:1996he} computed the first (differentially) rotating
BSs and found that these solutions have an integer ratio of angular
momentum to particle number. The structure of spinning BS models
has been studied extensively over the years
\cite{Ryan:1996nk,Yoshida:1997jq,Yoshida:1997qf,Yoshida:1997nd,Schunck:1999pm,Kleihaus:2005me,Kleihaus:2007vk,Kleihaus:2011sx,Collodel:2017biu}.
The quantised nature of the angular momentum also applies to Proca
and {\it Dirac} (spin $\tfrac{1}{2}$) stars \cite{Herdeiro:2019mbz},
but numerical studies of the formation of rotating stars have
revealed a striking difference between the scalar and vector case:
while collapsing scalar fields shed all their angular momentum
through an axisymmetric instability, the collapse of vector fields
results in spinning Proca stars with no indication of an instability
\cite{Sanchis-Gual:2019ljs,DiGiovanni:2020ror}.  This observation
is supported by analytic calculations \cite{Dmitriev:2021utv}, but
the instability may be quenched by self-interaction terms in the
potential function or in the Newtonian limit \cite{Siemonsen:2020hcg}.
For further reviews of the structure and dynamics of single BSs,
we note the reviews
\cite{Mielke:1997re,Mielke:2000mh,Mundim:2010hi,Liebling:2012fv}.

The first simulations of BS binaries have considered the head-on
collision of configurations with phase differences between the
constituent stars or opposite frequencies \cite{Palenzuela:2006wp};
see also \cite{Choptuik:2009ww,Bezares:2017mzk}.  The phase or
frequency differences manifest themselves most pronouncedly in the
dynamics and GW emission at late times around merger. These collisions
result in either a BH, a non-rotating BS or a near-annihilation of
the scalar field in the case of opposite frequencies. BS binaries
with orbital angular momentum generate a GW signal qualitatively
similar to that of BH binaries during the inspiral phase, but exhibit
a much more complex structure around merger
\cite{Palenzuela:2007dm,Palenzuela:2017kcg}.  In agreement with the
above mentioned BS formation studies, the BS inspirals also seem
to avoid the formation of spinning BSs, although they may settle
down into single nonrotating BSs.

In spite of the rapid progress of this field, the computation of
GW templates for BSs still lags considerably behind that of BH
binaries, both in terms of precision and coverage of the parameter
space.  Clearly, the presence of the matter fields adds complexity
to this challenge, but also alleviates some of the difficulties
through the non-singular character of the BS spacetimes. The first
main goal of our study is to highlight the substantial risk of
obtaining spurious physical results due to the use of overly
simplistic initial data constructed by plain superposition of
single-BS spacetimes. Our second main goal is to demonstrate how
an astonishingly simple modification of the superposition procedure,
first identified by Helfer {\it et al.}~\cite{Helfer:2018vtq} for
oscillatons, overcomes most of the problems encountered with plain
superposition. We summarise our main findings as follows.
\begin{list}{\rm{\bf (\arabic{count})}}{\usecounter{count}
             \labelwidth0.5cm \leftmargin1.0cm \labelsep0.2cm \rightmargin0.0cm
             \parsep0.5ex plus0.2ex minus0.1ex \itemsep0ex plus0.2ex}
  \item An adjustment of the superposition procedure, given 
  by Eq.~(\ref{eq:superposplus}), results in a significant
  reduction of the constraint violations inherent to the initial data;
  see Fig.~\ref{fig:ham}.
  \item In the head-on collision of mini BS binaries with rather
  low compactness, we observe a significant drop of the radiated
  GW energy with increasing distance $d$ if we use plain superposition.
  This physically unexpected dependence on the initial separation
  levels off only for rather large $d\gtrsim 150\,M$, where $M$
  denotes the Arnowitt-Deser-Misner (ADM) mass \cite{Arnowitt:1962hi}.
  In contrast, the total radiated energy computed from the evolution
  of our adjusted initial data displays the expected behaviour over
  the entire studied range $75.5\,M\le d\le 176\,M$: a very mild
  increase in the radiated energy with $d$. In the limit of large
  $d \gtrsim 150\,M$, both types of simulations agree within numerical
  uncertainties; see upper panel in Fig.~\ref{fig:erad}.
  \item In collisions of highly compact BSs with solitonic potentials,
  the radiated energy is largely independent of the initial separations
  for both initial data types, but for plain superposition we
  consistently obtain $\sim 10\,\%$ more radiation than for the
  adjusted initial data; see bottom panel in Fig.~\ref{fig:erad}.
  Furthermore, we find plain superposition to result in a slightly
  faster infall.  The most dramatic difference, however, is the
  collapse into individual BHs of both BSs well before merger if
  we use plain superposition. No such collapse occurs if we use
  adjusted initial data. Rather, these lead to the expected
  near-constancy of the central scalar-field amplitude of the BSs
  throughout most of the infall; see Fig.~\ref{fig:soli_ampctr}.
  \item We have verified through evolutions of single boosted BSs
  that the premature collapse into a BH is closely related to the
  spurious metric perturbation (\ref{eq:metricpert}) that arises
  in the plain superposition procedure. Artificially adding the
  same perturbation to a single BS spacetime induces an unphysical
  collapse of the BS that is in qualitative and quantitative agreement
  with that observed in the binary evolution starting with plain
  superposition; see Fig.~\ref{fig:soli_ampctr}.
\end{list}
The detailed derivation of these results begins in Sec.~\ref{sec:formalism}
with a review of the formalism and the computational framework of
our BS simulations. We discuss in more detail in
Sec.~\ref{sec:superposition} the construction of initial data through
plain superposition and our modification of this method. In
Sec.~\ref{sec:results}, we compare the dynamics of head-on collisions
of mini BSs and highly compact solitonic BS binaries starting from
both types of initial data. We note the substantial differences in
the results thus obtained and argue why we regard the results
obtained with our modification to be correct within numerical
uncertainties.  We summarise our findings and discuss future
extensions of this work in Sec.~\ref{sec:conclusions}.

Throughout this work, we use units where the speed of light and
Planck's constant are set to unity, $c=\hbar=1$. We denote spacetime
indices by Greek letters running from 0 to 3 and spatial indices
by Latin indices running from 1 to 3.

%=============================================================================
\section{Formalism}
\label{sec:formalism}

%=============================================================================
\subsection{Action and covariant field equations}
The action for a complex scalar field $\varphi$ minimally coupled
to gravity is given by
\begin{equation}
S = \int\sqrt{-g}\left\{
  \frac{1}{16\pi G}R - \frac{1}{2}\left[ 
  g^{\mu \nu}\nabla_{\mu}\bar{\varphi}\nabla_{\nu}\varphi
  + V(\varphi)\right] \right\} \du^4 x\,,
  \label{eq:action}
\end{equation}
where $g_{\alpha\beta}$ denotes the spacetime metric and $R$ the
Ricci scalar associated with this metric. The characteristics of
the resulting BS models depend on the scalar potential $V(\varphi)$;
in this work, we consider {\it mini boson stars} and {\it solitonic
boson stars}, obtained respectively for the potential functions
\begin{equation}
  V_{\rm min} = \mu^2 |\varphi|^2\,,~~~~~~~~~~
  V_{\rm sol} = \mu^2 |\varphi|^2\left(1-2\frac{|\varphi|^2}{\sigma_0^2}
  \right)^2\,.
  \label{eq:pot}
\end{equation}
Here, $\mu$ denotes the mass of the scalar field and $\sigma_0$
describes the self-interaction in the solitonic potential which can
result in highly compact stars \cite{Lee:1986ts}. Note that $V_{\rm
sol}\rightarrow V_{\rm min}$ in the limit $\sigma_0\rightarrow
\infty$.

Variation of the action (\ref{eq:action}) with respect to the metric
and the scalar field yield the Einstein and matter evolution equations
\begin{eqnarray}
  &&G_{\alpha\beta} = 8\pi G T_{\alpha\beta}
  = 8\pi G\left[
  \partial_{(\alpha}\bar{\varphi}\partial_{\beta)}\varphi
  -\frac{1}{2}g_{\alpha\beta}
  \big(
  g^{\mu\nu}\partial_{\mu}
  \bar{\varphi}\partial_{\nu}\varphi+ V(\varphi)
  \big)
  \right]\,, \label{eq:Einstein} \\
  &&\nabla^{\mu}\nabla_{\mu}\varphi = \varphi V'
  \defeq \varphi \frac{\du}{\du |\varphi|^2}
  V\,.
  \label{eq:Boxvarphi}
\end{eqnarray}
Readers who are mainly interested in the results of our work and/or
are familiar with the equations governing BS spacetimes may proceed
directly to Sec.~\ref{sec:superposition}.

%=============================================================================
\subsection{3+1 formulation}
For all simulations performed in this work, we employ the 3+1
spacetime split of ADM \cite{Arnowitt:1962hi} and York \cite{York1979};
see also \cite{Gourgoulhon:2007ue}. Here, the spacetime metric is
decomposed into the physical 3-metric $\gamma_{ij}$, the shift
vector $\beta^i$ and the lapse function $\alpha$ according to
\begin{equation}
  \du s^2 = g_{\alpha\beta}\du x^{\alpha}\du x^{\beta}
  = -\alpha^2 \du t^2 + \gamma_{mn}(\du x^m+\beta^m \du t)
  (\du x^n + \beta^n \du t)\,,
\end{equation}
where the level sets $x^0=t=\mathrm{const}$ represent three-dimensional
spatial hypersurfaces with timelike unit normal $n_{\mu}$.  Defining
the extrinsic curvature
\begin{equation}
  K_{ij} = -\frac{1}{2\alpha}(\partial_t \gamma_{ij}
  -\beta^m\partial_m \gamma_{ij} -\gamma_{im}\partial_j
  \beta^m-\gamma_{mj}\partial_i \beta^m)\,,
  \label{eq:Kij}
\end{equation}
the Einstein equations result in a first-order-in-time set of
differential equations for $\gamma_{ij}$ and $K_{ij}$ that is readily
converted into the conformal
Baumgarte-Shapiro-Shibata-Nakamura-Oohara-Kojima (BSSNOK) formulation
\cite{Baumgarte:1998te,Shibata:1995we,Nakamura:1987zz}.  More
specifically, we define
\begin{eqnarray}
  &&\chi = \gamma^{-1/3}\,,~~~~
  K=\gamma^{mn}K_{mn}\,,~~~~
  \tilde{\gamma}_{ij}=\chi \gamma_{ij}\,, \nonumber \\
  &&
  \tilde{A}_{ij}=\chi\left(K_{ij}-\frac{1}{3}\gamma_{ij}K\right)\,,~~~~
  \tilde{\Gamma}^i=\tilde{\gamma}^{mn}\tilde{\Gamma}^i_{mn}\,,
\end{eqnarray}
where $\gamma=\det \gamma_{ij}$, and $\tilde{\Gamma}^i_{mn}$ are
the Christoffel symbols associated with $\tilde{\gamma}_{ij}$.  The
Einstein equations are then given by (see for example Sec.~6 in
\cite{Cardoso:2014uka} for more details)
\begin{align}
  \partial_t \chi &= \beta^m\partial_m \chi
  + \frac{2}{3}\chi( \alpha K-\partial_m \beta^m)\,,
  \label{eq:chit} \\
  \partial_t \tilde{\gamma}_{ij} &=
  \beta^m\partial_m \tilde{\gamma}_{ij}
  + 2 \tilde{\gamma}_{m(i}\partial_{j)}\beta^m
  - \frac{2}{3}\tilde{\gamma}_{ij}\partial_m \beta^m
  -2\alpha \tilde{A}_{ij}\,, \\
  \partial_t K &=
  \beta^m\partial_m K
  -\chi \tilde{\gamma}^{mn}D_m D_n\alpha
  +\alpha \tilde{A}^{mn}\tilde{A}_{mn}
  +\frac{1}{3}\alpha K^2
  +4\pi G \alpha (S+\rho)\,, \\
  \partial_t \tilde{A}_{ij} &=
  \beta^m\partial_m \tilde{A}_{ij}
  +2\tilde{A}_{m(i}\partial_{j)}\beta^m
  -\frac{2}{3}\tilde{A}_{ij}\partial_m \beta^m
  +\alpha K \tilde{A}_{ij}
  -2\alpha\tilde{A}_{im}\tilde{A}^{m}{}_j
  \nonumber \\
  &~~
  +\chi (\alpha \mathcal{R}_{ij}-D_i D_j \alpha -8\pi G
        \alpha S_{ij})^{\rm TF}\,, \\
  \partial_t \tilde{\Gamma}^i &=
  \beta^m \partial_m \tilde{\Gamma}^i
  +\frac{2}{3}\tilde{\Gamma}^i\partial_m \beta^m
  -\tilde{\Gamma}^m\partial_m \beta^i
  +\tilde{\gamma}^{mn}\partial_m\partial_n \beta^i
  +\frac{1}{3}\tilde{\gamma}^{im}\partial_m\partial_n\beta^n
  \nonumber \\
  &~~~
  -\tilde{A}^{im}\left( 3\alpha \frac{\partial_m \chi}{\chi}
        +2\partial_m \alpha\right)
  +2\alpha\tilde{\Gamma}^i_{mn}\tilde{A}^{mn}
  -\frac{4}{3}\alpha\tilde{\gamma}^{im}\partial_m K
  -16\pi G \frac{\alpha}{\chi}j^i\,,
  \label{eq:Gammait}
\end{align}
where `TF' denotes the trace-free part and auxiliary expressions
are given by
\begin{align}
  \Gamma^i_{jk} &=
  \tilde{\Gamma}^i_{jk}
  -\frac{1}{2\chi}( \delta^i{}_k\partial_j \chi
        +\delta^i{}_j\partial_k \chi
        -\tilde{\gamma}_{jk}\tilde{\gamma}^{im}\partial_m \chi)
        \,, \nonumber\\
  \mathcal{R}_{ij} &= \tilde{R}_{ij}+ \mathcal{R}^{\chi}_{ij}
  \,, \nonumber\\
  \mathcal{R}^{\chi}_{ij} &=
  \frac{\tilde{\gamma}_{ij}}{2\chi}
  \left[
  \tilde{\gamma}^{mn}\tilde{D}_m\tilde{D}_n \chi
  -\frac{3}{2\chi}\tilde{\gamma}^{mn}\partial_m \chi\,\partial_n \chi
  \right]
  +\frac{1}{2\chi}
  \left(
  \tilde{D}_i \tilde{D}_j \chi
  -\frac{1}{2\chi}\partial_i \chi \,\partial_j \chi
  \right)
  \,, \nonumber\\
  \tilde{R}_{ij} &=
  -\frac{1}{2}\tilde{\gamma}^{mn}\partial_m \partial_n\tilde{\gamma}_{ij}
  +\tilde{\gamma}_{m(i}\partial_{j)}\tilde{\Gamma}^m
  +\tilde{\Gamma}^m\tilde{\Gamma}_{(ij)m}
  +\tilde{\gamma}^{mn}
  \left[
  2\tilde{\Gamma}^k_{m(i}\tilde{\Gamma}_{j)kn}
  +\tilde{\Gamma}^k_{im}\tilde{\Gamma}_{kjn}
  \right]\,,\nonumber\\
  D_i D_j \alpha &=
  \tilde{D}_i \tilde{D}_j \alpha
  + \frac{1}{\chi}\partial_{(i}\chi \partial_{j)}\alpha
  -\frac{1}{2\chi}\tilde{\gamma}_{ij}\tilde{\gamma}^{mn}
        \partial_m \chi \,\partial_n \alpha\,.
\end{align}
Here, $\tilde{D}$ and $\tilde{\mathcal{R}}$ denote the covariant
derivative and the Ricci tensor of the conformal metric
$\tilde{\gamma}_{ij}$, respectively.

The matter terms in Eqs.~(\ref{eq:chit})-(\ref{eq:Gammait}) are
defined by
\begin{equation}
  \rho = T_{\mu\nu}n^{\mu}n^{\nu}\,,~~~
  j_{\alpha} = -\bot^{\nu}{}_{\alpha} T_{\mu\nu} n^{\mu}\,,~~~
  S_{\alpha\beta} = \bot^{\mu}{}_{\alpha} \bot^{\nu}{}_{\beta}
        T_{\mu\nu}\,,~~~
  \bot^{\mu}{}_{\alpha}=\delta^{\mu}{}_{\alpha}+n^{\mu}n_{\alpha}\,.
\end{equation}
In adapted coordinates, we only need $\rho$ and the spatial components
$j^i$, $S_{ij}$ which are determined by the scalar field through
Eq.~(\ref{eq:Einstein}), Defining, in analogy to the extrinsic
curvature (\ref{eq:Kij}),
\begin{equation}
  \Pi = -\frac{1}{2\alpha}
  (
  \partial_t \varphi - \beta^m \partial_m \varphi
  )
  ~~~~~\Leftrightarrow~~~~~
  \partial_t \varphi = \beta^m\partial_m \varphi-2\alpha \Pi\,,
  \label{eq:Pi}
\end{equation}
we obtain
\begin{eqnarray}
  \rho &=&
  2\Pi \,\bar{\Pi}
  +\frac{1}{2}\partial^m \bar{\varphi}\,\partial_m\varphi
  +\frac{1}{2}V\,,~~~~~
  S+\rho = 8\bar{\Pi}\,\Pi-V\,, \nonumber \\
  j_i &=&
  \bar{\Pi}\partial_i \varphi
  +\Pi \partial_i \bar{\varphi}\,, \nonumber\\
  S_{ij} &=& \partial_{(i}\bar{\varphi}\partial_{j)}\varphi
  - \frac{1}{2}\gamma_{ij}
  \left(
  \gamma^{mn}\partial_m \bar{\varphi}\,\partial_n \varphi
  -4\bar{\Pi} \,\Pi
  +V
  \right)\,. \label{eqn:projectionofstressenergy}
\end{eqnarray}
The evolution of the scalar field according to Eq.~(\ref{eq:Boxvarphi})
in terms of our 3+1 variables is given by Eq.~(\ref{eq:Pi}) and
\begin{equation}
  \partial_t \Pi =
  \beta^m \partial_m \Pi
  + \alpha
  \left[
  \Pi K
  + \frac{1}{2}V'\varphi
  + \frac{1}{4} \tilde{\gamma}^{mn}
  \left(
  \partial_m \varphi\partial_n\chi
  -2\chi\tilde{D}_m\tilde{D}_n \varphi
  \right)
  \right]
  -\frac{1}{2} \chi\tilde{\gamma}^{mn}\partial_m\varphi
  \partial_n \alpha
  \,,
\end{equation}
where $V'=\du V/\du (|\varphi|^2)$

Finally, we evolve the gauge variables $\alpha$ and $\beta^i$ with
1+log slicing and the $\Gamma$-driver condition (the so-called
moving puncture conditions \cite{Campanelli:2005dd,Baker:2005vv}),
\begin{equation}
  \partial_t \alpha = \beta^m \partial_m \alpha
  -2\alpha K
  \,,~~~
  \partial_t \beta^i = \beta^m\partial_m \beta^i
  +\frac{3}{4}B^i
  \,,~~~
  \partial_t B^i = \beta^m \partial_m B^i
  +\partial_t \tilde{\Gamma}^i
  -\eta B^i\,,
\end{equation}
where $\eta$ is a constant we typically set to $M\eta\approx 1$ in
units of the ADM mass $M$.

Additionally to the evolution equations (\ref{eq:chit})-(\ref{eq:Gammait}),
the Einstein equations also imply four equations that do not contain
time derivatives, the Hamiltonian and momentum constraints
\begin{align}
  &\mathcal{H} \defeq \mathcal{R}+K^2-K^{mn}K_{mn} - 16\pi \rho = 0\,,
  \label{eq:ham} \\
  &\mathcal{M}_i\defeq D_i K - D_m K^m{}_i + 8\pi j_i = 0\,.
  \label{eq:mom}
\end{align}
While the constraints are preserved under time evolution in the
continuum limit, some level of violations is inevitable due to
numerical noise or imperfections of the initial data. We will return
to this point in more detail in Sec.~\ref{sec:superposition} below.

For the time evolutions discussed in Sec.~\ref{sec:results}, we
have implemented the equations of this section in the {\sc lean}
code \cite{Sperhake:2006cy} which is based on the {\sc cactus}
computational toolkit \cite{Allen:1999}. The equations are integrated
in time with the method of lines using the fourth-order Runge-Kutta
scheme with a Courant factor $1/4$ and fourth-order spatial
discretisation. Mesh refinement is provided by {\sc carpet}
\cite{Schnetter:2003rb} in the form of ``moving boxes'' and we
compute apparent horizons with {\sc AHFinderDirect}
\cite{Thornburg:1995cp,Thornburg:2003sf}.

%=============================================================================
\subsection{Stationary boson stars and initial data}
The initial data for our time evolution are based on single stationary
BS solutions in spherical symmetry.  Using spherical polar coordinates,
areal radius and polar slicing, the line element can be written as
\begin{equation}
  \du s^2 =
  -e^{2\Phi} \du t^2
  + \left(1-\frac{2m}{r}\right)^{-1} \du r^2
  + r^2
  (
  \du \theta^2
  + \sin^2\theta \du \phi^2
  )\,.
  \label{eq:ds2sym}
\end{equation}
where $\Phi$ and $m$ are functions of $r$ only.  It turns out
convenient to express the complex scalar field in terms of amplitude
and frequency,
\begin{equation}
  \varphi(t,r) =
  A(r)
  e^{\iu \omega t}\,,~~~~~\omega = \mathrm{const}\in \mathbb{R}\,.
\end{equation}
At this point, our configurations are characterised by two scales,
the scalar mass $\mu$ and the gravitational constant\footnote{Or,
equivalently, the Planck mass $M_{\rm Pl}=\sqrt{\hbar c/G}=1/\sqrt{G}$
for $\hbar=c=1$.} $G$.  In the following, we absorb $\mu$ and $G$
by rescaling all dimensional variables according to
\begin{equation}
  \hat{t}=\mu t\,,~~~
  \hat{r}=\mu r\,,~~~
  \hat{m}=\mu m\,,~~~
  \hat{A}=\sqrt{G} A\,,~~~
  \hat{\omega}=\omega/\mu\,;
  \label{eq:rescaling}
\end{equation}
note that $\mu$ has the dimension of a frequency or wave number and
$\sqrt{G}$ is an inverse mass. Using the Planck mass $M_{\rm
Pl}=1/\sqrt{G}=1.221\times10^{19}\,{\rm GeV}$, we can restore SI
units from the dimensionless numerical variables according to
\begin{equation}
  r = \hat{r} \times
  \left(
  \frac{\mu}{1.937\times 10^{-10}\,{\rm eV}}
  \right)^{-1}\,{\rm km}\,,~~~
  \omega = \hat{\omega}\times
  \frac{\mu}{6.582\times 10^{-16}\,{\rm eV}}\,{\rm Hz}
  \,,~~~
  A=\hat{A}\,M_{\rm Pl}\,,
  \nonumber
\end{equation}
and likewise for other variables. The rescaled version of the
potential (\ref{eq:pot}) is given by
\begin{equation}
  \hat{V}_{\rm min} = \hat{A}^2\,,~~~~~~~~~~
  \hat{V}_{\rm sol} =
  \hat{A}^2
  \left(
  1-2\frac{\hat{A}^2}{\hat{\sigma}_0^2}
  \right)^2~~\text{with}~~\hat{\sigma}_0=\sqrt{G}\sigma_0\,.
  \label{eq:potentials}
\end{equation}
In terms of the rescaled variables, the Einstein-Klein-Gordon
equations in spherical symmetry become
\begin{align}
  \partial_{\hat{r}}\Phi &=
  \frac{\hat{m}}{\hat{r}(\hat{r}-2\hat{m})}
  +
  2\pi \hat{r}
  \left(
  \hat{\eta}^2
  +\hat{\omega}^2e^{-2\Phi}\hat{A}^2
  -\hat{V}
  \right)\,, \label{eq:Phir} \\
  \partial_{\hat{r}}\hat{m} &=
  2\pi \hat{r}^2
  \left(
  \hat{\eta}^2
  +\hat{\omega}^2e^{-2\Phi}\hat{A}^2
  +\hat{V}
  \right)\,,
  \\
  \partial_{\hat{r}}\hat{A} &=
  \left(
  1-\frac{2\hat{m}}{\hat{r}}
  \right)^{-1/2}
  \hat{\eta}\,, \\
  \partial_{\hat{r}} \hat{\eta} &=
  -2\frac{\hat{\eta}}{\hat{r}}
  -\hat{\eta}\partial_{\hat{r}}\Phi
  +\left(
  1-\frac{2\hat{m}}{\hat{r}}
  \right)^{-1/2}
  (\hat{V}'-\hat{\omega}^2e^{-2\Phi})\hat{A}~~~
  \text{with}~~
  \hat{V}'=\frac{\du \hat{V}}{\du (\hat{A})^2}\,.
  \label{eq:etar}
\end{align}
By regularity, we have the following boundary conditions at the
origin $\hat{r}=0$ and at infinity,
\begin{equation}
  \hat{A}(0) = \hat{A}_{\rm ctr} \in \mathbb{R}^+\,,~~~~~
  \hat{m}(0) = 0\,,~~~~~
  \hat{\eta}(0) = 0\,,~~~~~
  \Phi(\infty) = 0\,,~~~~~
  \hat{A}(\infty) = 0\,.
\end{equation}
This two-point-boundary-value problem has two free parameters, the
central amplitude $\hat{A}_{\rm ctr}$ and the frequency $\hat{\omega}$.
For a given value $\hat{A}_{\rm ctr}$, however, only a discrete
(albeit infinite) number of frequency values $\hat{\omega}$ will
result in models with $\hat{A}(\infty)=0$; all other frequencies
lead to an exponentially divergent scalar field as $r\rightarrow
\infty$. The ``correct'' frequencies are furthermore ordered by
$\hat{\omega}_n<\hat{\omega}_{n+1}$, where $n\ge 0$ is the number
of zero crossings of the scalar profile $\hat{A}(\hat{r})$; $n=0$
corresponds to the ground state and $n>0$ to the $n^{\rm th}$ excited
state \cite{Balakrishna:1997ej}.  Finding the frequency for a regular
star for user-specified $\hat{A}_{\rm ctr}$ and $n$ is the key
challenge in computing BS models. We obtain these solutions through
a shooting algorithm, starting with the integration of
Eqs.~(\ref{eq:Phir})-(\ref{eq:etar}) outwards for $\hat{A}(0)=\hat{A}_{\rm
ctr}$ specified, $\Phi(0)=1$, and our ``initial guess'' $\hat{\omega}=1$.
Depending on the number of zero crossings in this initial-guess
model, we repeat the calculation by increasing or decreasing
$\hat{\omega}$ by one order of magnitude until we have obtained an
upper and a lower limit for $\hat{\omega}$. Through iterative
bisection, we then rapidly converge to the correct frequency. Because
we can only determine $\hat{\omega}$ to double precision, we often
find it necessary to capture the scalar field behaviour at large
radius by matching to its asymptotic behaviour
\begin{equation}
  \varphi \sim \frac{1}{\hat{r}}
  \exp\left( {-\sqrt{1-\hat{\omega}^2e^{-2\Phi}}}\right)\,,
\end{equation}
outside a user-specified radius $\hat{r}_{\rm match}$. Finally, we
can use an additive constant to shift the function $\Phi(\hat{r})$
to match its outer boundary condition. In practice, we impose this
condition in the form of the Schwarzschild relation
$e^{2\Phi}=(1-2\hat{m}/\hat{r})$ at the outer edge of our grid; in
vacuum this is exact even at finite radius, and we can safely ignore
the scalar field at this point thanks to its exponential falloff.
\begin{figure}[b]
    \centering
    \includegraphics[width=250pt]{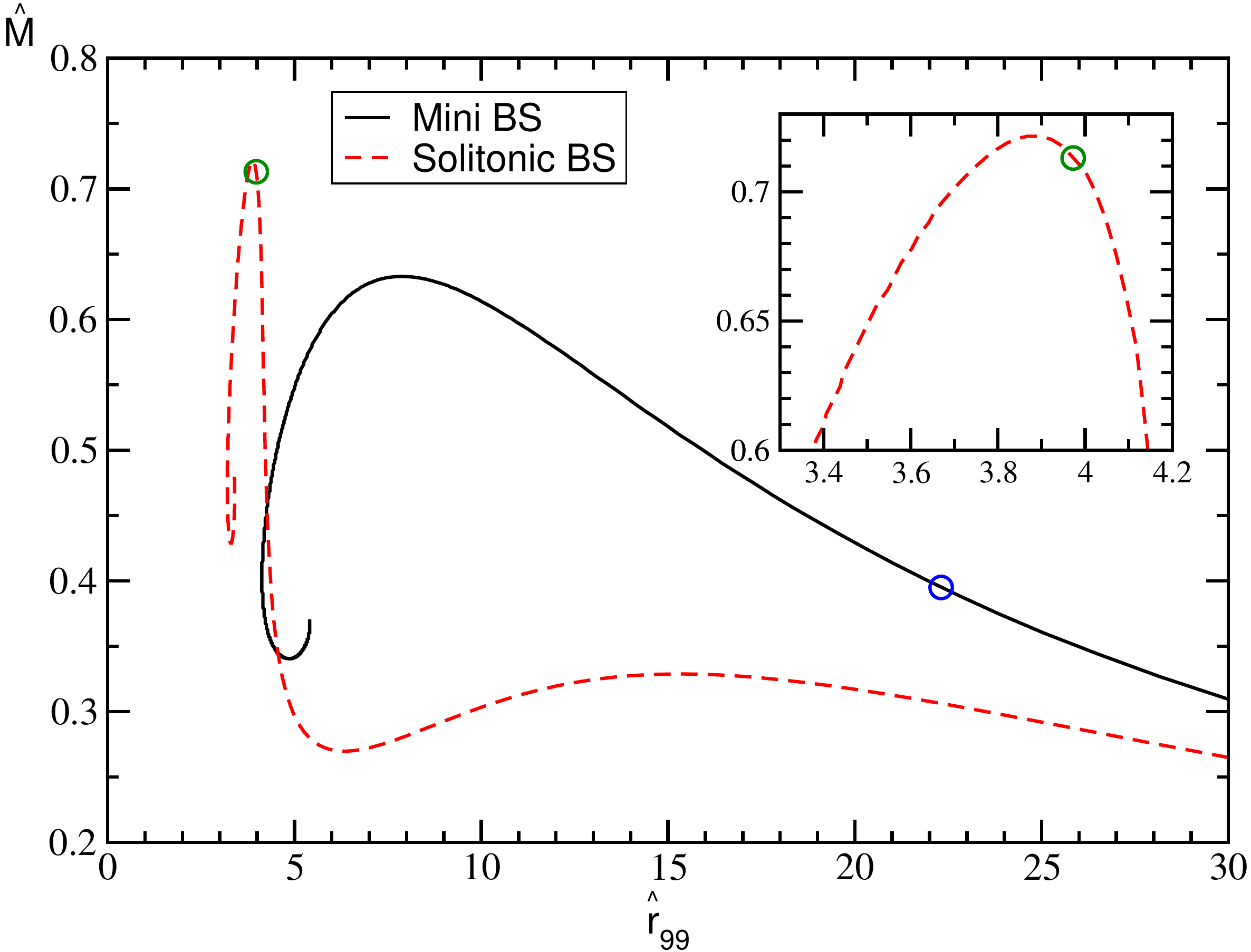}
    \caption{One parameter families of mini BSs (black solid) with
    a non-interacting potential $\hat{V}_{\rm min}$ and solitonic
    BSs (red dashed) with potential $\hat{V}_{\rm sol}$ and
    $\hat{\sigma}_0=0.2$ as given in Eq.~(\ref{eq:potentials}). In
    Sec.~\ref{sec:results} we simulate head-on collisions of two
    specific models marked by the circles and with parameters listed
    in Table \ref{tab:models}.
    }
    \label{fig:statBS}
\end{figure}

For a given potential, the solutions computed with this method form
a one-parameter family characterised by the central scalar field
amplitude $\hat{A}_0$. In Fig.~\ref{fig:statBS} we display two such
families for the potentials (\ref{eq:potentials}) with $\hat{\sigma}_0=0.2$
in the mass-radius diagram using the areal radius $\hat{r}_{99}$
containing $99\,\%$ of the BS's total mass. In that figure, we have
also marked by circles two specific models, one mini BS and one
solitonic BS, which we use in the head-on collisions in
Sec.~\ref{sec:results} below.  We have chosen these two models to
represent one highly compact and one rather squishy BS; note that
both models are located to the right of the maximal $\hat{M}(\hat{r})$
and, hence, stable stars.  Their parameters and properties are
summarised in Table \ref{tab:models}.
\begin{table}
    \centering
    \begin{tabular}{l|cccccc}
    \hline
    Model & $\sqrt{G}A_{\rm ctr}$ & $\sqrt{G}\sigma_0$ & $\mu M_{\rm BS}$ & $\omega/\mu$ & $\mu r_{99}$ & $\max\frac{m(r)}{r}$  \\
    \hline
    mini & 0.0124 & $\infty$ & $0.395$ & $0.971$ & $22.31$ & $0.0249$ \\
    soli & 0.17 & $0.2$ & $0.713$ & $0.439$ & $3.98$ & $0.222$ \\
    \hline
    \end{tabular}
    \caption{Parameters of the two single, spherically symmetric
    ground state BS models employed for our simulations of head-on
    collisions. Up to the rescaling with the scalar mass $\mu$,
    each BS is determined by the central amplitude $A_{\rm ctr}$
    of the scalar field and the potential parameter $\sigma_0$ of
    Eq.~(\ref{eq:pot}). The mass $M_{\rm BS}$ of the boson star,
    the scalar field frequency $\omega$, the areal radius $r_{99}$
    containing $99\,\%$ of the total mass $M_{\rm BS}$ and the
    compactness, defined here as the maximal ratio of the mass
    function to radius, represent the main features of the stellar
    model.}
    \label{tab:models}
\end{table}

The formalism discussed so far provides us with BS solutions in
radial gauge and polar slicing.  In order to reduce the degree of
gauge adjustment in our moving puncture time evolutions, however,
we prefer using conformally flat BS models in isotropic gauge. In
isotropic coordinates, the line element of a spherically symmetric
spacetime has the form
\begin{equation}
  \mu^2 \du s^2 = -e^{2\Phi}\du \hat{t}^2+\psi^4 (\du \hat{R}^2
        + \hat{R}^2\du \Omega^2)\,,
  \label{eq:ds2iso}
\end{equation}
where $\du \Omega^2=\du\theta^2+\sin^2\theta\,\du \phi^2$. Comparing
this with the polar-areal line element (\ref{eq:ds2sym}), we obtain
two conditions,
\begin{equation}
  \psi^4 \hat{R}^2 = \hat{r}^2\,,~~~~~~~~
  \psi^4\du \hat{R}^2=
  X^2 \du \hat{r}^2~~~\text{with}~~~
  X=\left(1-\frac{2\hat{m}}{\hat{r}}\right)^{-1/2}\,.
\end{equation}
In terms of the new variable $f(\hat{r})=\hat{R}/\hat{r}$, we obtain
the differential equation
\begin{equation}
  \frac{\du f}{\du \hat{r}} = \frac{f}{\hat{r}}(X-1)\,,
\end{equation}
which we integrate outwards by assuming $\hat{R}\propto \hat{r}$
near $\hat{r}=0$. The integrated solution can be rescaled by a
constant factor to ensure that at large radii -- where the scalar
field has dropped to a negligible level -- we recover the Schwarzschild
value $\psi = 1+\frac{\hat{m}}{2\hat{R}}$, in accordance with
Birkhoff's theorem.  Bearing in mind that $\psi^4\hat{R}^2=\hat{r}^2$,
this directly leads to the outer boundary condition
\begin{equation}
  \hat{R}_{\rm ob} =
  \frac{\hat{r}_{\rm ob}-\hat{m}_{\rm ob}}{2}
  \left[
  1+
  \sqrt{1-\frac{\hat{m}_{\rm ob}^2}{(\hat{r}_{\rm ob}-\hat{m}_{\rm ob})^2}}
  \right]\,,
\end{equation}
end, hence, the overall scaling factor applied to the function
$\hat{R}(\hat{r})$.

In isotropic coordinates, the resulting spacetime metric is trivially
converted from spherical to Cartesian coordinates
$\hat{x}^i=(\hat{x},\hat{y},\hat{z})$ using $\du \hat{R}^2 +
\hat{R}^2\du \Omega^2=\du \hat{x}^2+\du \hat{y}^2+\du \hat{z}^2$,
so that
\begin{equation}
  \mu^2 \du s^2 = -e^{2\Phi} \du \hat{t}^2 + \psi^4 \delta_{ij}
        \du \hat{x}^i \,\du \hat{x}^j \,.
\end{equation}
For convenience, will drop the caret on the rescaled coordinates
and variables from now on and implicitly assume that they represent
dimensionless quantities to be converted into dimensional form
according to Eq.~(\ref{eq:rescaling}).

%=============================================================================
\subsection{Boosted boson stars}
The single BS solutions can be converted into boosted stars through
a straightforward Lorentz transformation. For this purpose, we
denote the star's rest frame by $\mathcal{O}$ with Cartesian 3+1
coordinates $x^{\alpha}=(t,\,x^k)$ and consider a second frame
$\tilde{\mathcal{O}}$ with Cartesian 3+1 coordinates
$\tilde{x}^{\tilde{\alpha}}=(\tilde{t},\,\tilde{x}^{\tilde{k}})$
that moves relative to $\mathcal{O}$ with constant velocity $v^i$.
These two frames are related by the transformation
\begin{equation}
  \Lambda^{\tilde{\alpha}}{}_{\mu} = \left( \begin{array}{c|c}
        \gamma & -\gamma v_j \\
        \hline
        -\gamma v^i & \delta^i{}_j+(\gamma-1)\frac{v^i v_j}{|\vec{v}|^2}
  \end{array} \right)~~~~\Leftrightarrow~~~~
  \Lambda^{\mu}{}_{\tilde{\alpha}} = \left( \begin{array}{c|c}
        \gamma & \gamma v_j \\
        \hline
        \gamma v^i & \delta^i{}_j+(\gamma-1)\frac{v^i v_j}{|\vec{v}|^2}
  \end{array} \right)
  \,.
  \nonumber
\end{equation}
Starting with the isotropic rest-frame metric (\ref{eq:ds2iso}) and
the complex scalar field $\varphi(t,R)=A(R)e^{\iu \omega t +
\vartheta_0}$ with $R=\sqrt{\delta_{mn}x^mx^n}$, we obtain a general
boosted model in terms of the 3+1 variables in Cartesian coordinates
$\tilde{x}^{\tilde{k}}$ as follows.
\begin{list}{\rm{\bf (\arabic{count})}}{\usecounter{count}
             \labelwidth0.5cm \leftmargin1.0cm \labelsep0.2cm \rightmargin0.0cm
             \parsep0.5ex plus0.2ex minus0.1ex \itemsep0ex plus0.2ex}
  \item A straightforward calculation leads to the first derivatives
  of the metric, its inverse and the scalar field in Cartesian
  coordinates $x^i$ in the rest frame,
  \begin{eqnarray}
    \partial_t g_{\mu\nu}=\partial_t g^{\mu\nu}=0\,,~~~&&
    \partial_t \varphi_R = -\omega \varphi_I\,,~~~
    \partial_t \varphi_I = \omega \varphi_R\,,
    \nonumber \\[5pt]
    \partial_i g_{00} = -2e^{2\Phi}\frac{\du \Phi}{\du R} \frac{x^i}{R}\,,
    ~~~~~&&
    \partial_i g^{00} = 2e^{-2\Phi} \frac{\du \Phi}{\du R}\frac{x^i}{R}\,,
    \nonumber \\[5pt]
    \partial_i g_{kk} = 4\psi^3 \frac{\du \psi}{\du R} \frac{x^i}{R}\,,
    && \partial_i g^{kk} = -4\psi^{-5} \frac{\du \psi}{\du R}
          \frac{x^i}{R}\,,
    \nonumber \\[5pt]
    \partial_i \varphi_R = \frac{\eta}{f} \cos(\omega t + \phi_0)\,
        \frac{x^i}{R}\,,~~~~&&
  \partial_i \varphi_I = \frac{\eta}{f} \sin(\omega t + \phi_0)\,
        \frac{x^i}{R}\,,
  \end{eqnarray}
  where $\varphi_R$ and $\varphi_I$ are the real and imaginary part
  of the scalar field, and
  \begin{equation}
    \frac{\du \psi}{\du R} = -\frac{1}{2}
        \frac{X-1}{X} \frac{\psi}{R}\,,~~~~~
        \displaystyle \frac{\du \Phi}{\du R}=
        \frac{X^2-1}{2XR}+\frac{2\pi R}{f^2}X(\eta^2+\omega^2
        e^{-2\Phi}A^2-V)\,.
  \end{equation}
  \item We Lorentz transform the spacetime metric, the scalar field
  and their derivatives to the boosted frame $\tilde{\mathcal{O}}$
  according to
  \begin{gather}
    \tilde{g}_{\tilde{\alpha}\tilde{\beta}}=
      \Lambda^{\mu}{}_{\tilde{\alpha}}\Lambda^{\nu}{}_{\tilde{\beta}}g_{\mu\nu}
      \,,~~~
    \tilde{g}^{\tilde{\alpha}\tilde{\beta}}=
    \Lambda^{\tilde{\alpha}}{}_{\mu}\Lambda^{\tilde{\beta}}{}_{\nu}g^{\mu\nu}
    \,,~~~
    \partial_{\tilde{\gamma}}\tilde{g}_{\tilde{\alpha}\tilde{\beta}}=
    \Lambda^{\lambda}{}_{\tilde{\gamma}}\Lambda^{\mu}{}_{\tilde{\alpha}}
    \Lambda^{\nu}{}_{\tilde{\beta}}\partial_{\lambda}g_{\mu\nu}
    \,, \nonumber\\
     \tilde{\varphi}(\tilde{x}^{\alpha}) = \varphi(x^{\mu})\,,~~~
    \partial_{\tilde{\alpha}}\tilde{\varphi}=
    \Lambda^{\mu}{}_{\tilde{\alpha}} \partial_{\mu}\varphi\,.
  \end{gather}
  \item We construct the 3+1 variables in the boosted frame from these
  quantities according to
  \begin{align}
    &\tilde{\alpha} = \left(-\tilde{g}^{\tilde{0}\tilde{0}}\right)^{-1/2}
    \,,~~~
    \tilde{\beta}_{\tilde{k}}=\tilde{g}_{\tilde{0}\tilde{k}}
    \,,~~~
    \tilde{\gamma}_{\tilde{k}\tilde{l}}=\tilde{g}_{\tilde{k}\tilde{l}}
    \,,~~~
    \tilde{\beta}^{\tilde{k}}=\tilde{\gamma}^{\tilde{k}\tilde{m}}
    \tilde{\beta}_{\tilde{m}}
    \,, \\
    & \tilde{K}_{\tilde{k}\tilde{l}}=
    -\frac{1}{2\tilde{\alpha}}
    \left(
    \partial_{\tilde{t}}\tilde{\gamma}_{\tilde{k}\tilde{l}}
    -\tilde{\beta}^{\tilde{m}}\partial_{\tilde{m}}\tilde{\gamma}_{\tilde{k}\tilde{l}}
    -\tilde{\gamma}_{\tilde{m}\tilde{l}}\partial_{\tilde{k}}\tilde{\beta}^{\tilde{m}}
    -\tilde{\gamma}_{\tilde{k}\tilde{m}}\partial_{\tilde{l}}\tilde{\beta}^{\tilde{m}}
    \right)\,, \nonumber \\
    &\tilde{\Pi} = -\frac{1}{2\tilde{\alpha}}
    \left(
    \partial_{\tilde{t}}\tilde{\varphi}
    - \tilde{\beta}^{\tilde{m}} \partial_{\tilde{m}}\tilde{\varphi}
    \right)\,,
  \end{align}
  with
  \begin{equation}
    \partial_{\tilde{t}}\tilde{\gamma}_{\tilde{k}\tilde{l}}=
    \partial_{\tilde{0}}\tilde{g}_{\tilde{k}\tilde{l}}
    \,,~~~~~
    \tilde{\gamma}_{\tilde{k}\tilde{m}}
    \partial_{\tilde{l}}\tilde{\beta}^{\tilde{m}}
    =
    \partial_{\tilde{l}}\tilde{g}_{\tilde{0}\tilde{k}}
    - \tilde{\beta}^{\tilde{m}}
    \partial_{\tilde{l}}\tilde{g}_{\tilde{k}\tilde{m}}\,.
  \end{equation}
  \item In addition to these expressions, we need to bear in mind
  the coordinate transformation. The computational domain of our
  time evolution corresponds to the boosted frame $\tilde{\mathcal{O}}$.
  A point $\tilde{x}^{\tilde{\alpha}}=(\tilde{t},\,\tilde{x}^{\tilde{k}})$
  in that domain therefore has rest-frame coordinates
  \begin{equation}
     (t,\,x^k) = x^{\mu} = \Lambda^{\mu}{}_{\tilde{\alpha}}
        \tilde{x}^{\tilde{\alpha}}\,.
  \end{equation}
  It is at $(t,\,x^k)$, where we need to evaluate the rest frame
  variables $\Phi(R)$, $X(R)$, the scalar field $\varphi(t,R)$ and
  their derivatives. In particular, different points on our initial
  hypersurface $\tilde{t}=0$ will in general correspond to different
  times $t$ in the rest frame.
\end{list}
%

%=============================================================================
\section{Boson-star binary initial data}
\label{sec:superposition}
The single BS models constructed according to the procedure of the
previous section are exact solutions of the Einstein equations,
affected only by a numerical error that we can control by increasing
the resolution, the size of the computational domain and the degree
of precision of the floating point variable type employed. The
construction of binary initial data is conceptually more challenging
due to the non-linear character of the Einstein equations; the
superposition of two individual solutions will, in general, not
constitute a new solution. Instead, such a superposition incurs
some violation of the constraint equations (\ref{eq:ham}),
(\ref{eq:mom}). The purpose of this section is to illustrate how
we can substantially reduce the degree of constraint violation with
a relatively simple adjustment in the superposition. Before introducing
this ``trick'', we first summarise the superposition as it is
commonly used in numerical simulations.

%=============================================================================
\subsection{Simple superposition of boson stars}
\label{sec:superpossimple}
The most common configuration involving more than one BS is a binary
system, and this is the scenario we will describe here. We note,
however, that the method generalises straightforwardly to any number
of stars.  Let us then consider two individual BS solutions with
their centres located at $x^i_{\rm A}$ and $x^i_{\rm B}$, velocities
$v_{\rm A}^i$ and $v_{\rm B}^i$.  The two BS spacetimes are described
by the 3+1 (ADM) variables $\gamma_{ij}^{\rm A}$, $\alpha_{\rm A}$,
$\beta_{\rm A}^i$ and $K_{ij}^{\rm A}$, the scalar field variables
$\varphi_{\rm A}$ and $\Pi_{\rm A}$, and likewise for star B. We
can construct from these individual solutions an approximation for
a binary BS system via the pointwise superposition
\begin{align}
  &\gamma_{ij} =
  \gamma_{ij}^{\rm A}
  +\gamma_{ij}^{\rm B}
  -\delta_{ij}\,,~~~~~~
  K_{ij} = \gamma_{m(i}
  \left[
  K^{\rm A}_{j)n}
  \gamma_{\rm A}^{nm}
  +K^{\rm B}_{j)n}
  \gamma_{\rm B}^{nm}
  \right]\,,
  \nonumber \\
  &\varphi = \varphi_{\rm A} + \varphi_{\rm B}\,,
  \hspace{0.85cm}~~~~~~~~~~
  \Pi = \Pi_{\rm A} + \Pi_{\rm B}\,.
  \label{eq:superpossimple}
\end{align}
One could similarly construct a superposition for the lapse $\alpha$
and shift vector $\beta^i$, but their values do not affect the
physical content of the initial hypersurface.  In our simulations
we instead initialise them by $\alpha=\sqrt{\chi}$ and $\beta^i=0$.

A simple superposition approach along the lines of
Eq.~(\ref{eq:superpossimple}) has been used in numerous studies of
BS as well as BH binaries including higher-dimensional BHs
\cite{Palenzuela:2006wp,Palenzuela:2007dm,Shibata:2008rq,Okawa:2011fv,Palenzuela:2017kcg,Sperhake:2019oaw}.
For BHs and higher-dimensional spacetimes in particular, this
leading-order approximation has proved remarkably successful and
in some limits a simple superposition is exact, such as infinite
initial separation, in Brill-Lindquist initial data for non-boosted
\setcounter{footnote}{6} BHs\footnote{Note that for Brill-Lindquist
one superposes the conformal factor $\psi$ rather than $\psi^4$ as
in the method discussed here.} \cite{Brill:1963yv} or in the
superposition of Aichelburg-Sexl shockwaves \cite{Aichelburg:1970dh}
for head-on collisions of BHs at the speed of light.  It has been
noted in Helfer {\em et al.}~\cite{Helfer:2018vtq}, however, that
this simple construction can result in spurious low-frequency
amplitude modulations in the time evolution of binary oscillatons
(real-scalar-field cousins of BSs); cf.~their Fig.~7.  Furthermore,
they have proposed a straightforward remedy that essentially
eliminates this spurious modulation. As we will see in the next
section, the repercussions of the {\it simple superposition} according
to Eqs.~(\ref{eq:superpossimple}) can be even more dramatic for BS
binaries, but they can be cured in the same way as in the oscillaton
case.  We note in this context that BSs may be more vulnerable to
superposition artefacts near their centres due to the lack of a
horizon and its potentially protective character in the superposition
of BHs.

The key problem of the construction (\ref{eq:superpossimple}) is
the equation for the spatial metric $\gamma_{ij}$. This is best
illustrated by considering the centre $x_{\rm A}^i$ of star A.  In
the limit of infinite separation, the metric field of its companion
star B becomes $\gamma_{ij}^{\rm B}\rightarrow \delta_{ij}$.  This
is, of course, precisely the contribution we subtract in the third
term on the right-hand-side and all would be well.  In practice,
however, the BSs start from initial positions $x_{\rm A}^i$ and
$x_{\rm B}^i$ with finite separation $d=||x_{\rm A}^i-x_{\rm B}^i||$
and we consequently perturb the metric at star A's centre by
\begin{equation}
  \delta \gamma_{ij} = \gamma_{ij}^{\rm B}(x^i_{\rm A})
  -\delta_{ij}
  \label{eq:metricpert}
\end{equation}
away from its equilibrium value $\gamma_{ij}^{\rm A}(x_{\rm A}^i)$.
This metric perturbation can be interpreted as a distortion of the
volume element $\sqrt{\gamma}$ at the centre of star A. More
specifically, the volume element at star A's centre is enhanced by
$\mathcal{O}(1)\,\%$ for initial separations $\mathcal{O}(100)\,M$
and likewise for the centre of star B (by symmetry); see appendix
A of Ref.~\cite{Helfer:2018vtq} for more details.\footnote{Due to
the slow decay of this effect $\propto 1/\sqrt{d}$ \cite{Helfer:2018vtq},
a simple cure in terms of using larger $d$ is often not practical.}
The energy density $\rho$, on the other hand, is barely altered by
the presence of the other star, because of the exponential fall-off
of the scalar field. The leading-order error therefore consists in
a small excess mass that has been added to each BS's central region.
We graphically illustrate this effect in the upper half of
\begin{figure}[t]
    \centering
    \includegraphics[width=350pt]{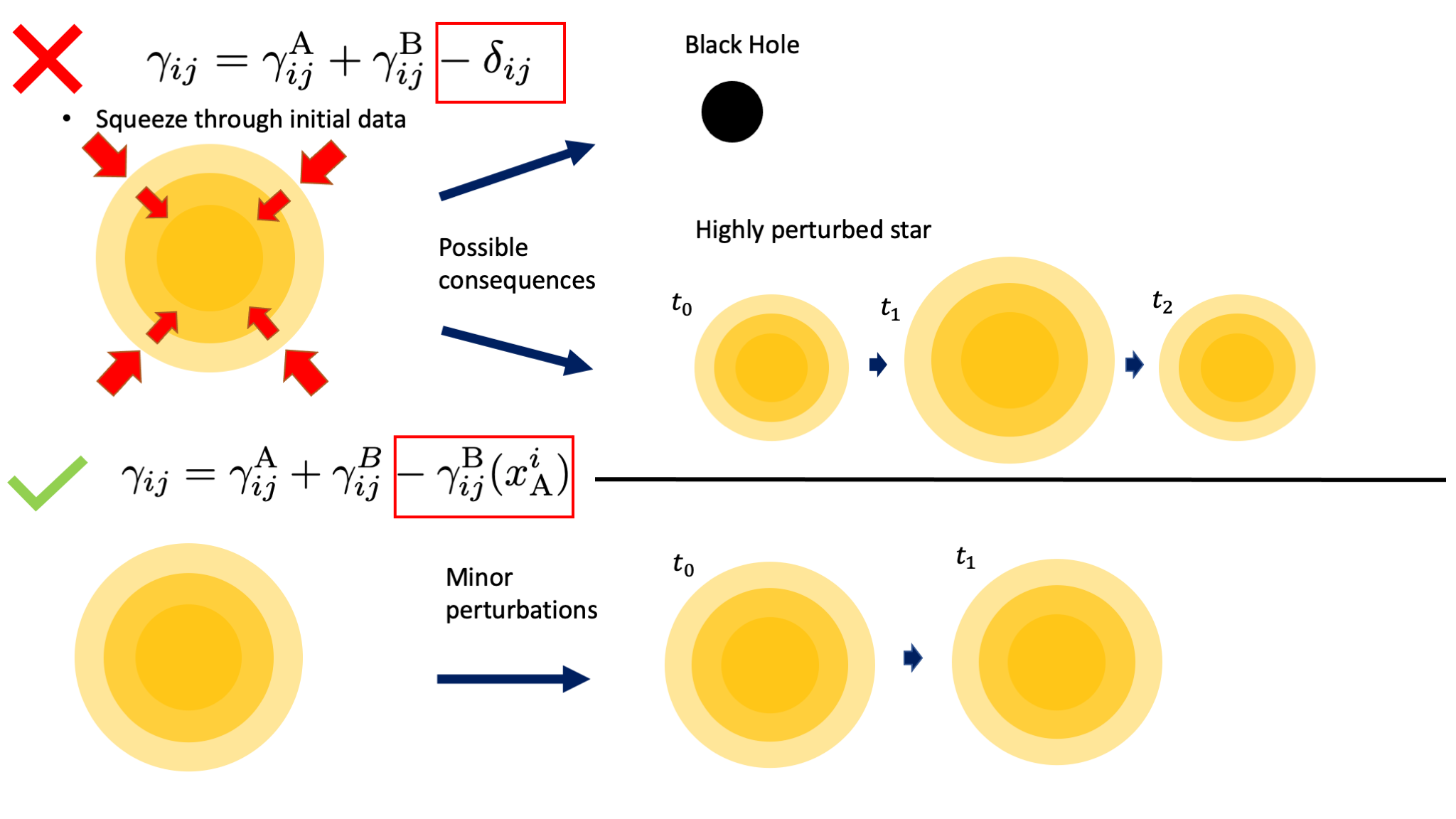}
    \caption{
    Graphical illustration of the spurious dynamics that may be
    introduced by the simple superposition procedure
    (\ref{eq:superpossimple}). {\it Upper panel}: The spurious
    increase in the volume element mimics a squeezing of the stellar
    core that effects a pulsation of the star or may even trigger
    gravitational collapse to a BH.  {\it Lower panel}: No such
    squeezing occurs with the adjusted superposition
    (\ref{eq:superposplus}), and the binary evolution starts with
    approximately unperturbed stars.
    }
    \label{fig:Overview}
\end{figure}
Fig.~\ref{fig:Overview} together with some of the possible consequences.
As we will see, this qualitative interpretation is fully borne out
by the phenomenology we observe in the binaries' time evolutions.

Finally, we would like to emphasise that, while evaluating the
constraint violations is in general a good rule of thumb to check
whether the field configuration is a solution of the system, it
does \emph{not} inform one whether it is \emph{the intended} solution;
a system with some constraint violation may have drifted closer to
a different, unintended solution. In the present case, in addition
to the  increased constraint violation, the constructed BS solutions
possess significant excitations. Thus, while applying a constraint
damping system like conformal Z4 \cite{Bernuzzi:2009ex,Alic:2011gg}
may eventually drive the system to a solution, it may no longer be
what was originally intended to be the initial condition of an
unexcited BS star.

%=============================================================================
\subsection{Improved superposition}
The problem of the simple superposition is encapsulated by
Eq.~(\ref{eq:metricpert}) and the resulting deviation of the volume
elements at the stars' centres away from their equilibrium values.
At the same time, the equation presents us with a concrete recipe
to mitigate this error: we merely need to replace in the simple
superposition (\ref{eq:superpossimple}) the first relation
$\gamma_{ij}=\gamma_{ij}^{\rm A}+\gamma_{ij}^B -\delta_{ij}$ by
\begin{equation}
  \boxed{
  \gamma_{ij}=\gamma_{ij}^{\rm A}+\gamma_{ij}^{\rm B}
  -\gamma_{ij}^{\rm B}(x^i_{\rm A})
  =\gamma_{ij}^{\rm A}+\gamma_{ij}^{\rm B}
  -\gamma_{ij}^{\rm A}(x^i_{\rm B})\,.}
  \label{eq:superposplus}
\end{equation}
The two expressions on the right-hand side are indeed equal thanks
to the symmetry of our binary: its constituents have equal mass,
no spin and their velocity components satisfy $v_{\rm A}^i v_{\rm
A}^j=v_{\rm B}^i v_{\rm B}^j$ for all $i,\,j=1,\,2,\,3$ in the
centre-of-mass frame.  Equation (\ref{eq:superposplus}) manifestly
ensures that at positions $x_{\rm A}^i$ and $x_{\rm B}^i$ we now
recover the respective star's equilibrium metric and, hence, volume
element.  We graphically illustrate this improvement in the bottom
panel of Fig.~\ref{fig:Overview}.

A minor complication arises from the fact that the resulting spatial
metric does not asymptote towards $\delta_{ij}$ as $R\rightarrow
\infty$. We accordingly impose outgoing Sommerfeld boundary conditions
on the asymptotic background metric $2\delta_{ij}-\gamma_{ij}^{\rm
A}(x^i_{\rm B})$; in a set of test runs, however, we find this
correction to result in very small changes well below the simulation's
discretisation errors.

Finally, we note that the leading-order correction to the superposition
as written in Eq.~(\ref{eq:superposplus}) does not work for asymmetric
configurations with unequal masses or spins. Generalising the method
to arbitrary binaries requires the subtraction of a spatially varying
term rather than a constant $\gamma_{ij}^{\rm B}(x^i_{\rm
A})=\gamma_{ij}^{\rm A}(x^i_{\rm B})$ or $\delta_{ij}$. Such a
generalisation may consist, for example, of a weighted sum of the
terms $\gamma_{ij}^{\rm A}(x^i_{\rm B})$ and $\gamma_{ij}^{\rm
B}(x^i_{\rm A})$. Leaving this generalisation for future work, we
will focus on equal-mass systems in the remainder of this study and
explore the degree of improvement achieved with
Eq.~(\ref{eq:superposplus}).

%=============================================================================
\section{Models and results}
\label{sec:results}
For our analysis of the two types of superposed initial data, we
will now discuss time evolutions of binary BS head-on collisions.
A head-on collision is characterised by the two individual BS models
and three further parameters, the initial separation in units of
the ADM mass, $d/M$, and the initial velocities $v_{\rm A}$ and
$v_{\rm B}$ of the BSs. We perform all our simulations in the
centre-of-mass frame, so that for equal-mass binaries, $v_{\rm
A}=-v_{\rm B}\invdefeq v$.  One additional parameter arises from
the type of superposition used for the initial data construction:
we either use the ``plain'' superposition of Eq.~(\ref{eq:superpossimple})
or the ``adjusted'' method (\ref{eq:superposplus}).

For all our simulations, we set $v=0.1$; this value allows us to
cover a wide range of initial separations without the simulations
becoming prohibitively long.
\begin{table}[t]
    \centering
    \caption{
    The four types of BS binary head-on collisions simulated in
    this study. The individual BSs A and B are given either by the
    mini or solitonic model of Table \ref{tab:models}, and start
    with initial velocity $v$ directed towards each other. The
    initial data is constructed either by plain superposition
    (\ref{eq:superpossimple}) or by adjusting the superposed data
    according to Eq.~(\ref{eq:superposplus}).  For each type of
    binary, we perform five collisions with initial separations $d$
    listed in the final column.
    }
    \begin{tabular}{r|ccccc}
    \hline
    Label & star A & star B & $v$ & initial data & $d/M$ \\
    \hline
    {\tt mini} & mini & mini & $0.1$ & plain &
    75.5,~101,~126,~151,~176 \\
    {\tt +mini}& mini & mini & $0.1$ & adjusted &
    75.5,~101,~126,~151,~176 \\
    {\tt soli} & soli & soli & $0.1$ & plain &
    16.7,~22.3,~27.9,~33.5,~39.1 \\
    {\tt +soli} & soli & soli & $0.1$ &
    adjusted &
    16.7,~22.3,~27.9,~33.5,~39.1 \\
    \hline
    \end{tabular}
    \label{tab:hods}
\end{table}
The BS binary configurations summarised in Table \ref{tab:hods}
then result in four sequences of head-on collisions labelled {\tt
mini}, {\tt +mini}, {\tt soli} and {\tt +soli}, depending in the
nature of the constituent BSs and the superposition method. For
each sequence, we vary the BSs initial separation $d$ to estimate
the dependence of the outcome on $d$. First, however, we test our
interpretation of the improved superposition (\ref{eq:superposplus})
by computing the level of constraint violations in the initial data.

%=============================================================================
\subsection{Initial constraint violations}
\begin{figure}
  \centering
  \includegraphics[width=300pt]{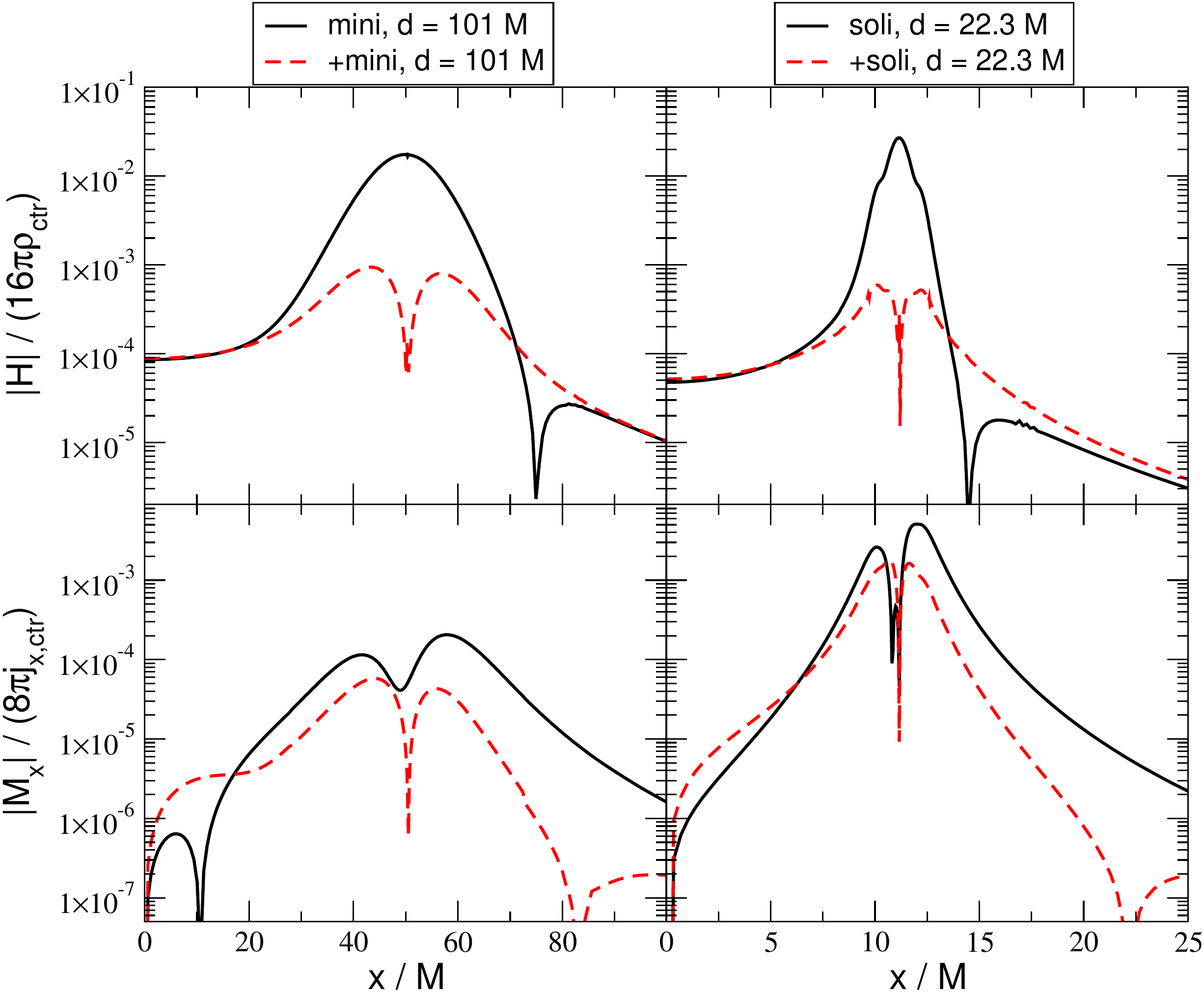}
  \caption{
  Upper row: The Hamiltonian constraint violation $\mathcal{H}$ --
  Eq.~(\ref{eq:ham}) -- normalised by the respective BS's central
  energy density $16\pi \rho_{\rm ctr}$ is plotted along the collision
  axis of the binary configurations {\tt mini}, {\tt +mini} with
  $d=101\,M$ (left) and {\tt soli}, {\tt +soli} with $d=22.3\,M$
  (right).  The degree of violations is substantially reduced in
  the BS interior by using the improved superposition
  (\ref{eq:superposplus}) for {\tt +mini} and {\tt +soli} relative
  to their plain counterparts; the maxima of $\mathcal{H}$ have
  dropped by over an order of magnitude in both cases.  Bottom row:
  The same analysis for the momentum constraint $\mathcal{M}_x$
  normalised by the central BS's momentum density $8\pi j_x$. Here
  the improvement is less dramatic, but still yields a reduction
  by a factor of a few in the BS core.
  }
  \label{fig:ham}
\end{figure}

As discussed in Sec.~\ref{sec:superpossimple} and in Appendix A of
Ref.~\cite{Helfer:2018vtq}, the main shortcoming of the plain
superposition procedure consists in the distortion of the volume
element near the individual BSs' centres and the resulting perturbation
of the mass-energy inside the stars away from their equilibrium
values. If this interpretation is correct, we would expect this
effect to manifest itself in an elevated level of violation of the
Hamiltonian constraint (\ref{eq:ham}) which relates the energy
density to the spacetime curvature. Put the other way round, we
would expect our improved method (\ref{eq:superposplus}) to reduce
the Hamiltonian constraint violation. This is indeed the case as
demonstrated in the upper panels of Fig.~\ref{fig:ham} where we
plot the Hamiltonian constraint violation of the initial data along
the collision axis for the configurations {\tt mini} and {\tt +mini}
with $d=101\,M$ and the configurations {\tt soli} and {\tt +soli}
with $d=22.3\,M$.

In the limit of zero boost velocity $v=0$, this effect is even
tractable through an analytic calculation which confirms that the
improved superposition (\ref{eq:superposplus}) ensures $\mathcal{H}=0$
at the BS's centres in isotropic coordinate; see \ref{sec:hamanalytic}
for more details.

Our adjustment (\ref{eq:superposplus}) also leads to a reduction
of the momentum constraint violations of the initial data, although
the effect is less dramatic here.  The bottom panels of Fig.~\ref{fig:ham}
display the momentum constraint $\mathcal{M}_x$ of Eq.~(\ref{eq:mom})
along the collision axis normalised by the momentum density $8\pi
j_x$; we see a reduction by a factor of a few over large parts of
the BS interior for the modified data {\tt +mini} and {\tt +soli}.

The overall degree of initial constraint violations is rather small
in all cases, well below $0.1\,\%$ for our adjusted data. These
data should therefore also provide a significantly improved initial
guess for a full constraint solving procedure. We leave such an
analysis for future work and in the remainder of the work explore
the impact of the adjustment (\ref{eq:superposplus}) on the physical
results obtained from the initial data's time evolutions.

%=============================================================================
\subsection{Convergence and numerical uncertainties}
In order to put any differences in the time evolutions into context,
we need to understand the uncertainties inherent to our numerical
simulations. For this purpose, we have studied the convergence of
the GW radiation generated by the head-on collisions of mini and
solitonic BSs.

\begin{figure}[t]
    \centering
    \includegraphics[width=250pt]{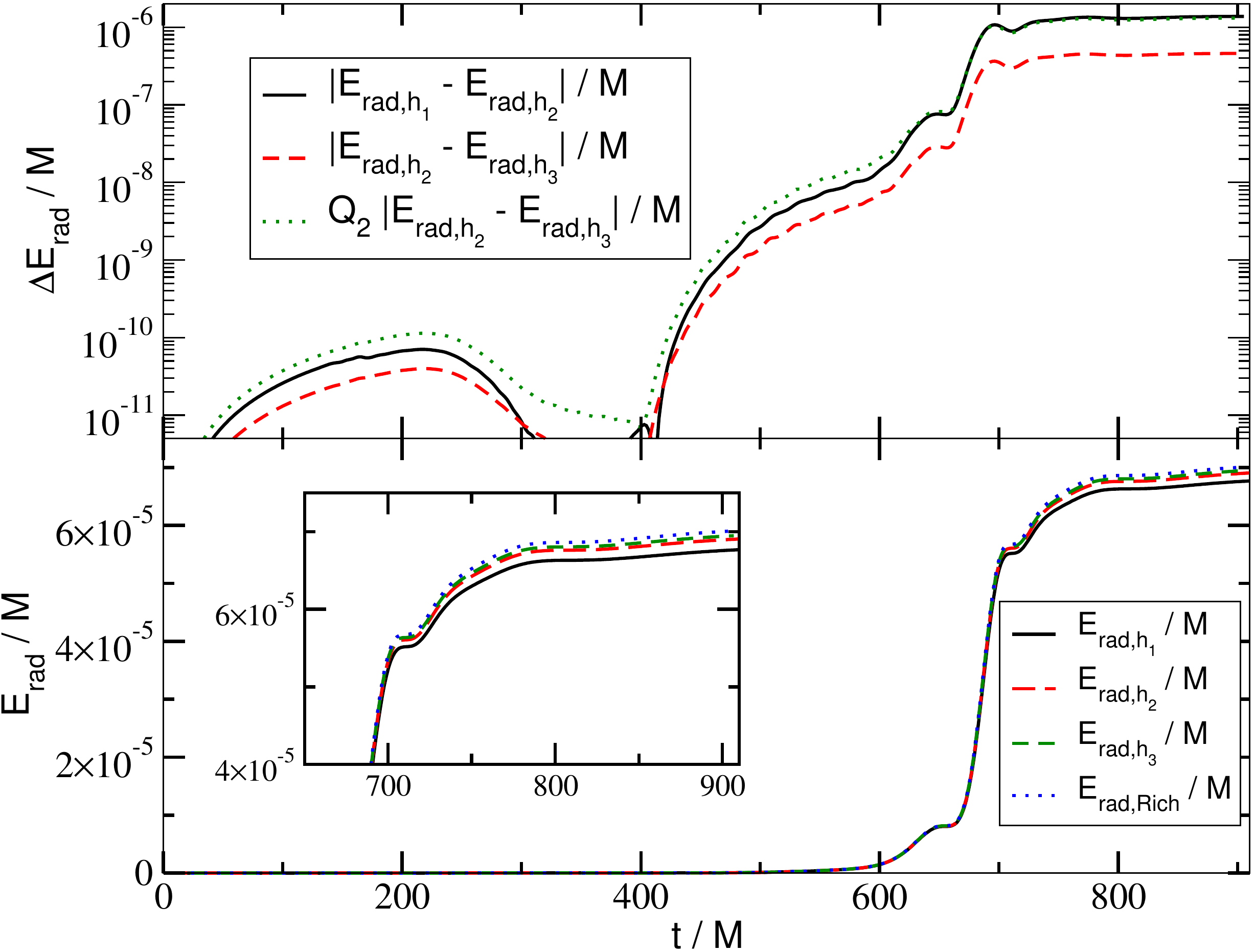}
    \caption{
    Convergence analysis for the GW energy extracted at $R_{\rm
    ex}=252\,M$ from the head-on collision {\tt +mini} of Table
    \ref{tab:models} with $d=101\,M$. For the resolutions $h_1=M/6.35$,
    $h_2=M/9.53$ and $h_3=12.70$ (on the innermost refinement level),
    we obtain convergence close to second order (upper panel). The
    numerical error, obtained by comparing our results with the
    second-order Richardson extrapolated values (bottom panel), is
    $0.9\,\%$ ($1.6\,\%$, $3.6\,\%$) for our high (medium, coarse)
    resolutions.
    }
    \label{fig:conv_tmBS_Erad}
\end{figure}

Figure \ref{fig:conv_tmBS_Erad} displays the convergence of the
radiated energy $E_{\rm rad}$ as a function of time for the {\tt
+mini} configuration with $d=101\,M$ of Table \ref{tab:models}
obtained for grid resolutions $h_1=M/6.35$, $h_2=M/9.53$ and
$h_3=M/12.70$ on the innermost refinement level and corresponding
grid spacings on the other levels. The functions $E_{\rm rad}(t)$
and their differences are shown in the bottom and top panel,
respectively, of Fig.~\ref{fig:conv_tmBS_Erad} together with an
amplification of the high-resolution differences by the factor
$Q_2=2.86$ for second-order convergence. The observation of
second-order convergence is compatible with the second-order
ingredients of the {\sc Lean} code, prolongation in time and the
outgoing radiation boundary conditions. We believe that this dominance
is mainly due to the smooth behaviour of the BS centre as compared
with the case of black holes \cite{Husa:2007hp}. By using the
second-order Richardson extrapolated result, we determine the
discretisation error of our energy estimates as $0.9\,\%$ for $h_3$
which is the resolution employed for all remaining mini BS collisions.
We have performed the same convergence analysis for the plain-superposition
counterpart {\tt mini} and for the dominant $(\ell,m)=(2,0)$ multipole
of the Newman-Penrose scalar of both configurations and obtained
the same convergence and very similar relative errors.

\begin{figure}[t]
    \centering
    \includegraphics[width=250pt]{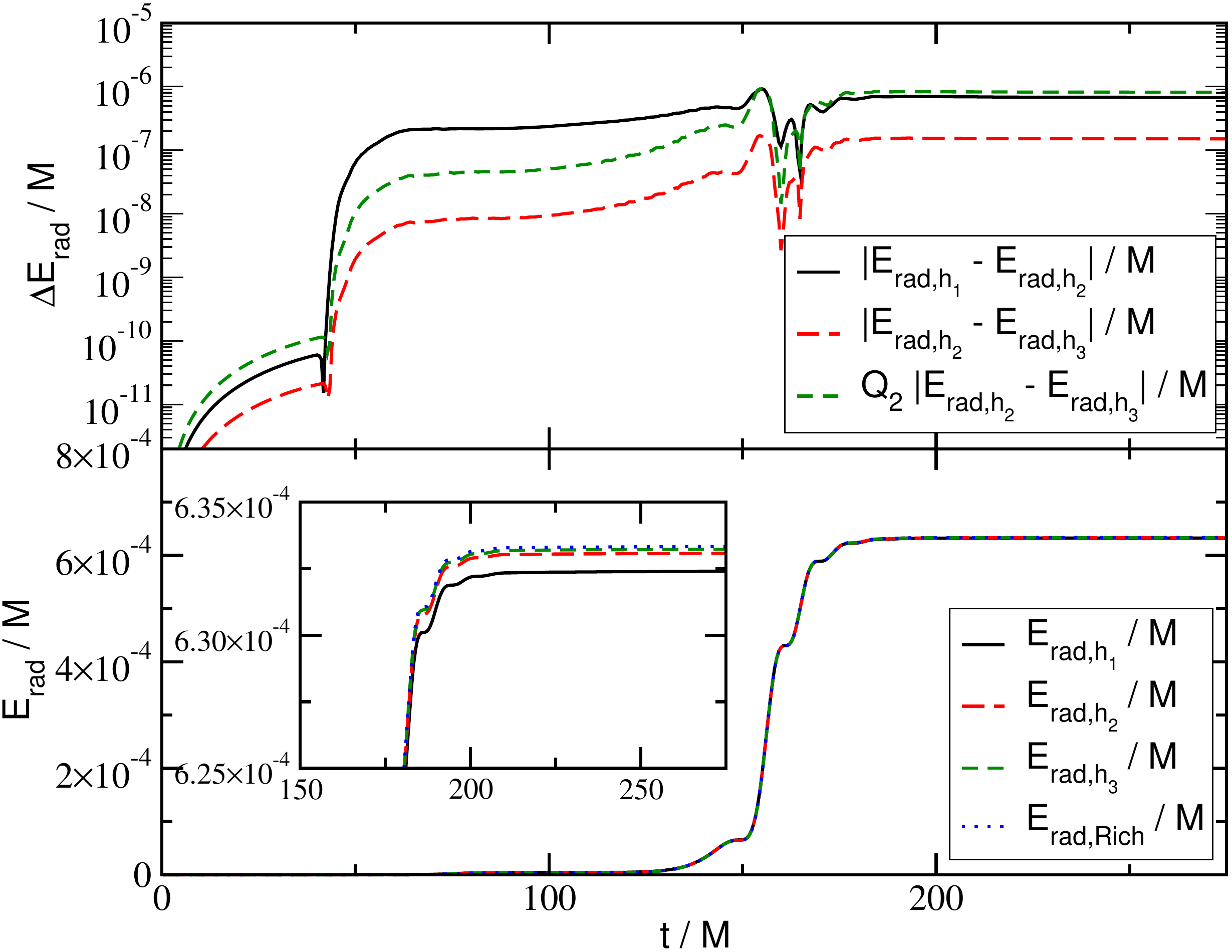}
    \caption{
    Convergence analysis as in Fig.~\ref{fig:conv_tmBS_Erad} but
    for the configuration {\tt +soli} of Table \ref{tab:models}
    with $d=22.3\,M$ and resolutions $h_1=M/22.9$, $h_2=M/45.9$ and
    $h_3=M/68.8$.  The numerical error, obtained by comparing our
    results with the second-order Richardson extrapolated values
    (bottom panel), is $0.03\,\%$ ($0.07\,\%$, $0.6\,\%$) for our
    high (medium, coarse) resolutions.
    }
    \label{fig:conv_tsBS_Erad}
\end{figure}

In Fig.~\ref{fig:conv_tsBS_Erad}, we show the same convergence
analysis for the solitonic collision {\tt +soli} with $d=22.3\,M$
and resolutions $h_1=M/22.9$, $h_2=M/45.9$, $h_3=M/68.8$. We observe
second-order convergence during merger and ringdown and slightly
higher convergence in the earlier infall phase.
For the uncertainty estimate we conservatively use the second-order
Richardson extrapolated result and obtain a discretisation error
of about $0.07\,\%$ for our medium resolution $h_2$ which is the
value we employ in our solitonic production runs. Again, we have
repeated this analysis for the plain {\tt soli} counterpart and the
$(2,0)$ GW multipole observing the same order of convergence and
similar uncertainties. Our error estimate for the solitonic
configurations is rather small in comparison to the mini BS collisions
and we cannot entirely rule out a fortuitous cancellation of errors
in our simulations.  From this point on, we therefore use a
conservative discretisation error estimate of $1\,\%$ for all our
BS simulations.

A second source of uncertainty in our results is due to the extraction
of the GW signal at finite radii rather than $\mathcal{I}^+$. We
determine this error by extracting the signal at multiple radii,
fitting the resulting data by the series expansion $f=f_0+f_1/r$,
and comparing the result at our outermost extraction radius with
the limit $f_0$. This procedure results in errors in $E_{\rm rad}$
ranging between $0.5\,\%$ and $3\,\%$. With the upper range, we
arrive at a conservative total error budget for discretisation and
extraction of about $4\,\%$.  As a final test, we have repeated the
{\tt mini} and {\tt +mini} collisions for $d=101\,M$ with the
independent {\sc GRChombo} code \cite{Clough:2015sqa,Radia:2021}
using the CCZ4 formulation \cite{Alic:2011gg} and obtain the same
results within $\approx 1.5\,\%$.  Bearing in mind these tests and
a $4\,\%$ error budget, we next study the dynamics of the BS head-on
collisions with and without our adjustment of the initial data.

%=============================================================================
\subsection{Radiated gravitational-wave energy}
For our first test, we compute the total radiated GW energy for all
our head-on collisions focusing in particular on its dependence on
the initial separation $d$ of the BS centres. In this estimate we
exclude any spurious or ``junk'' radiation content of the initial
data by starting the integration at $t=R_{\rm ex}+40\,M$. Unless
specified otherwise, all our results are extracted at $R_{\rm
ex}=300\,M$ for mini BS collisions and $R_{\rm ex}=84\,M$ for the
solitonic binaries.
\begin{figure}
    \centering
    \includegraphics[width=250pt]{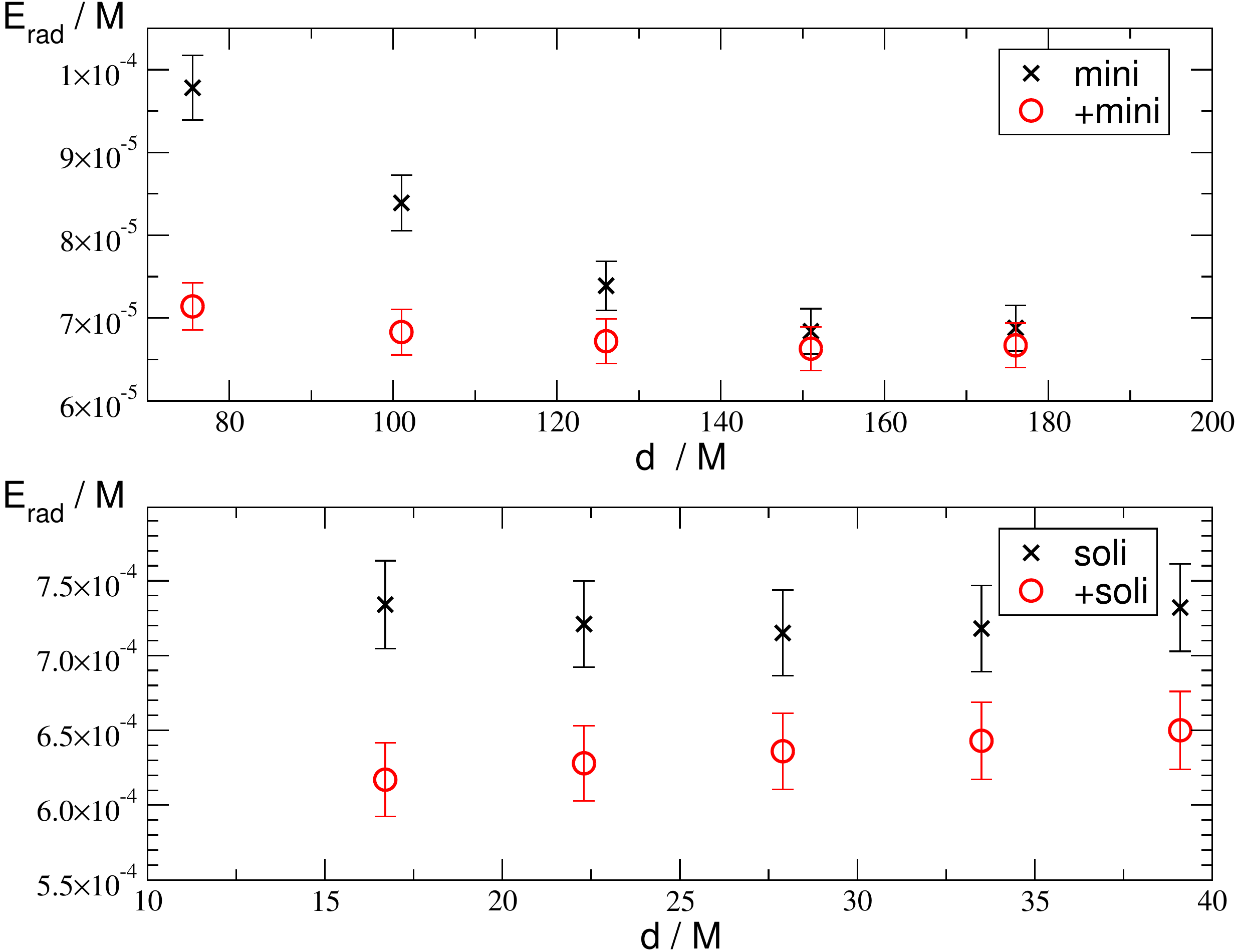}
    \caption{
    The GW energy $E_{\rm rad}$ generated in the head-on collision
    of mini (upper panel) and solitonic (lower panel) BS binaries
    starting with initial separation $d$ and velocity $v=0.1$ towards
    each other.  For comparison, a non-spinning, equal-mass BH
    binary colliding head-on with the same boost velocity $v=0.1$
    radiates $E_{\rm rad}=6.0\times 10^{-4}\,M$ \cite{Sperhake:2019oaw}.
    } 
    \label{fig:erad}
\end{figure}

The main effect of increasing the initial separation is a reduction
of the (negative) binding energy of the binary and a corresponding
increase of the collision velocity around merger. In the large $d$
limit, however, this effect becomes negligible. For the comparatively
large initial separations chosen in our collisions, we would therefore
expect the function $E_{\rm rad}$ to be approximately constant,
possibly showing a mild increase with $d$. The mini BS collisions
shown as black $\times$ symbols in the upper panel of Fig.~\ref{fig:erad}
exhibit a rather different behaviour: the radiated energy rapidly
decreases with $d$ and only levels off for $d\gtrsim 150\,M$. We
have verified that the excess energy for smaller $d$ is not due to
an elevated level of junk radiation which consistently contribute
well below $0.1\,\%$ of $E_{\rm rad}$ in all our mini BS collisions
and has been excluded from the results of Fig.~\ref{fig:erad} anyway.
The {\tt +mini} BS collisions, in contrast, results in an approximately
constant $E_{\rm rad}$ with a total variation approximately at the
level of the numerical uncertainties. For $d\gtrsim 150\,M$, both
types of initial data yield compatible results, as is expected.
The key benefit of our adjusted initial data is that they provide
reliable results even for smaller initial separations suitable for
starting BS inspirals.

The discrepancy is less pronounced for the head-on collisions of
solitonic BS collisions; both types of initial data result in
approximately constant $E_{\rm rad}$. They differ, however, in the
predicted amount of radiation at a level that is significant compared
to the numerical uncertainties. As we will see below, this difference
is accompanied by drastic differences in the BS's dynamics during
the long infall period. We furthermore note that the mild but steady
increase obtained for the adjusted {\tt +soli} agrees better with
the physical expectations.

\begin{figure}
    \centering
    \includegraphics[width=250pt]{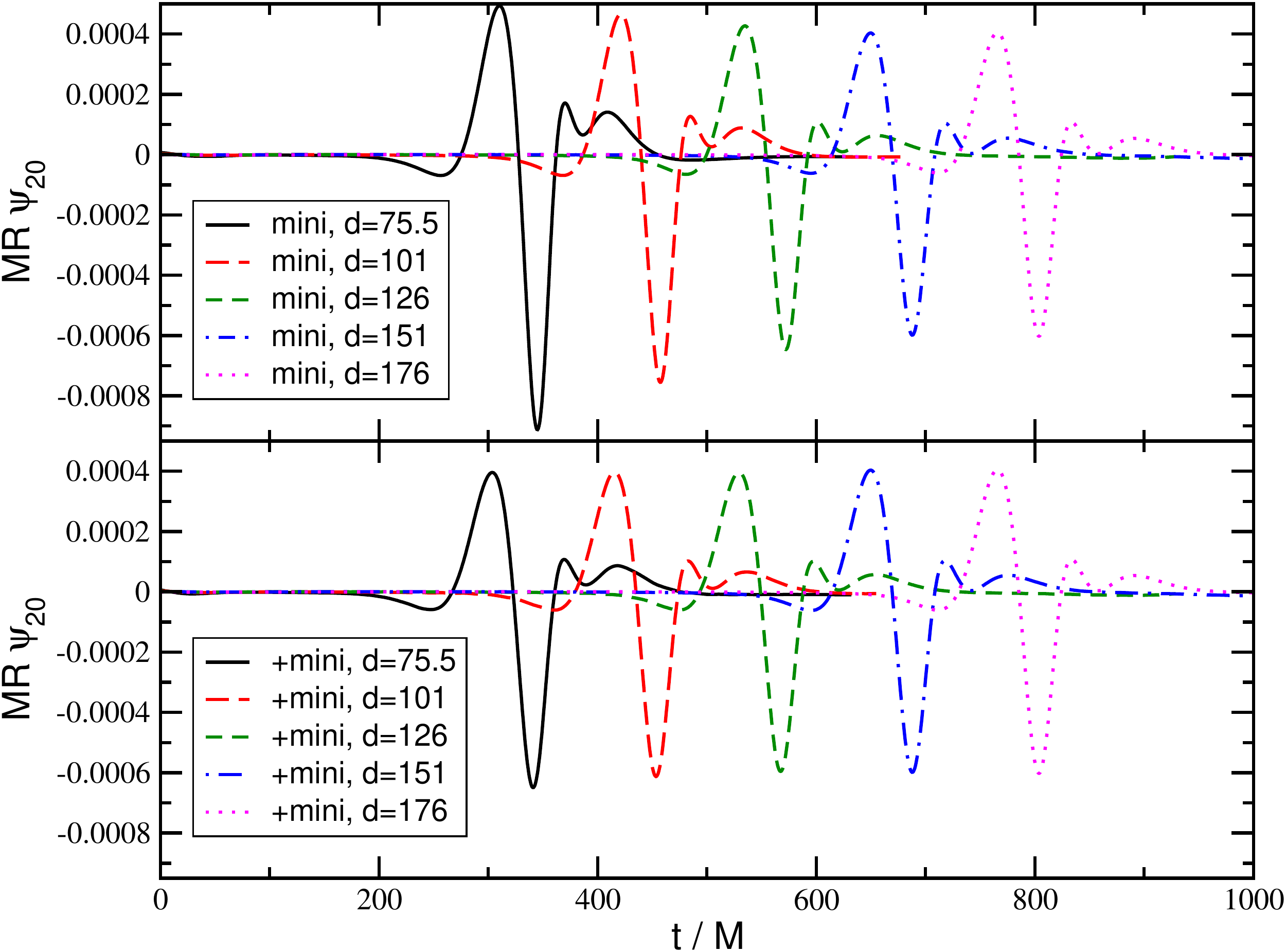}
    \caption{
    The $(2,0)$ mode of the Newman-Penrose scalar for the mini boson
    star collisions of Table \ref{tab:models}.
    }
    \label{fig:mini_psi20}
\end{figure}

\begin{figure}
    \centering
    \includegraphics[width=250pt]{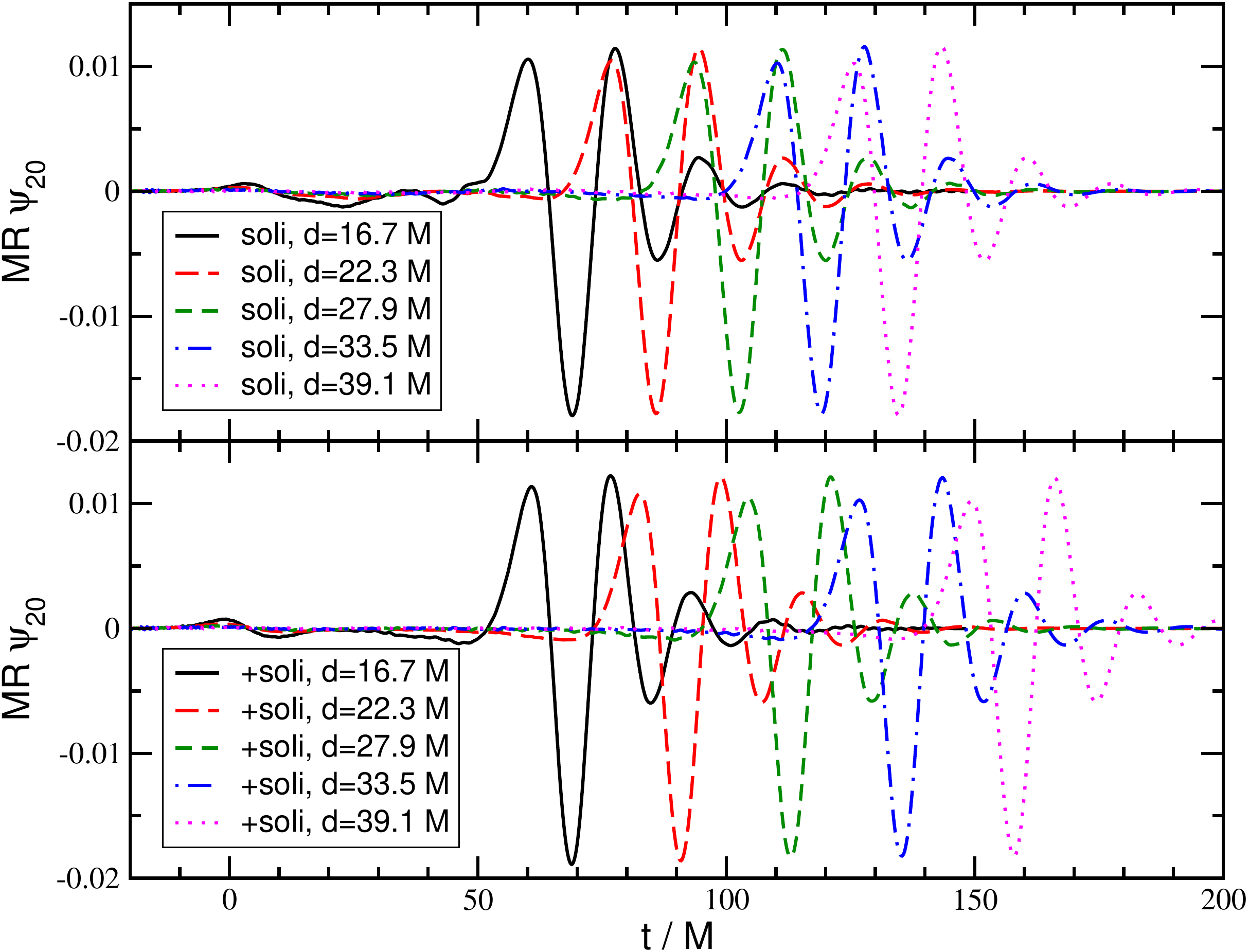}
    \caption{
    The $(2,0)$ mode of the Newman-Penrose scalar for the solitonic
    boson star collisions of Table \ref{tab:models}.
    }
    \label{fig:soli_psi20}
\end{figure}

The differences in the total radiated GW energy also manifest
themselves in different amplitudes of the $(2,0)$ multipole of the
Newman-Penrose scalar $\Psi_4$. This is displayed in
Figs.~\ref{fig:mini_psi20} and \ref{fig:soli_psi20} where we show
the GW modes for the mini and solitonic collisions, respectively.
The most prominent difference between the results for plain and
adjusted initial data is the significant variation of the amplitude
of the $(2,0)$ mode in the plain mini BS collisions in the upper
panel of Fig.~\ref{fig:mini_psi20}.  In contrast, the differences
in the amplitudes in Fig.~\ref{fig:soli_psi20} for the solitonic
collisions are very small. In fact, the differences in the radiated
energy of the {\tt soli} and {\tt +soli} collisions mostly arise
from a minor stretching of the signal for the {\tt soli} case; this
effect is barely perceptible in Fig.~\ref{fig:soli_psi20} but is
amplified by the integration in time when we calculate the energy.
Finally, we note the different times of arrival of the main pulses
in Fig.~\ref{fig:soli_psi20}; especially for larger initial separation,
the merger occurs earlier for the {\tt soli} configurations than
for their adjusted counterparts {\tt +soli}.  We will discuss this
effect together with the evolution of the scalar field amplitude
in the next subsection.

%

%=============================================================================
\subsection{Evolution of the scalar amplitude and gravitational collapse}
\begin{figure}
    \centering
    \includegraphics[width=250pt]{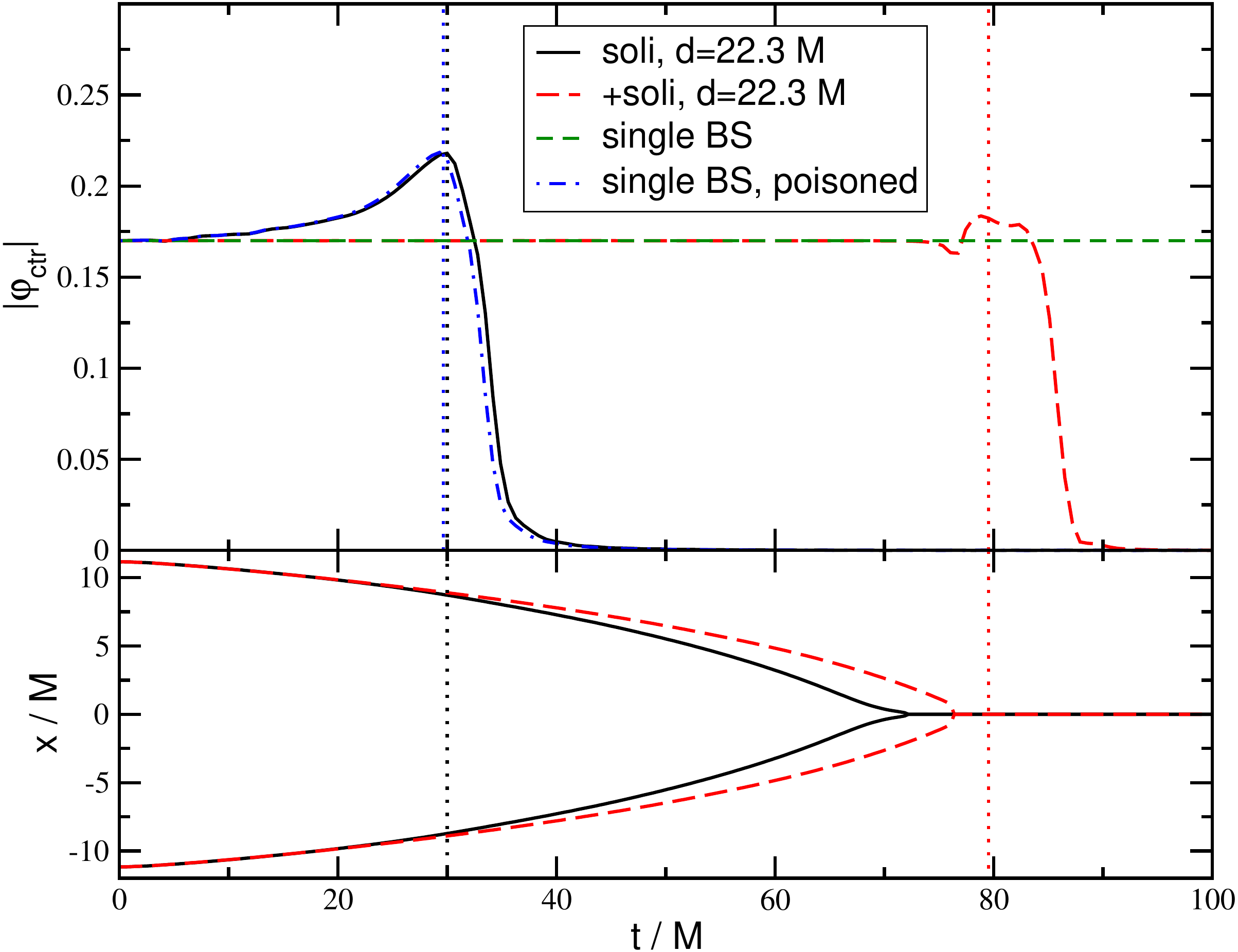}
    \caption{
    The central scalar-field amplitude $|\varphi_{\rm ctr}|$ as a
    function of time for one BS in the head-on collisions of solitonic
    BSs with distance $d=22.3\,M$ (black solid and red long-dashed)
    as well as a single BS spacetime with the same parameters (green
    dashed) and the same single BS spacetime ``poisoned'' with the
    metric perturbation (\ref{eq:metricpert}) that would arise in
    a simple superposition (see text for details). The dotted
    vertical lines mark the first location of an apparent horizon
    in the simulation of the same colour; as expected, no horizon
    ever forms in the evolution of the unpoisoned single BS.  In
    the bottom panel, we show for reference the coordinate trajectories
    of the BS centres as obtained from locally Gauss-fitting the
    scalar profile. Around merger this procedure becomes inaccurate,
    so that the values around $t\approx 70\,M$ should be regarded
    as qualitative measures, only.
    }
    \label{fig:soli_ampctr}
\end{figure}

The adjustment (\ref{eq:superposplus}) in the superposition of
oscillatons was originally developed in Ref.~\cite{Helfer:2018vtq}
to reduce spurious modulations in the scalar field amplitude;
cf.~their Fig.~7. In our simulations, this effect manifests itself
most dramatically in the collisions of our solitonic BS configurations
{\tt soli} and {\tt +soli}.  From Fig.~\ref{fig:statBS}, we recall
that the single-BS constituents of these binaries are stable, but
highly compact stars, located fairly close to the instability
threshold. We would therefore expect them to be more sensitive to
spurious modulations in their central energy density. This is exactly
what we observe in all time evolutions of the {\tt soli} configurations
starting with plain-superposition initial data. As one example, we
show in Fig.~\ref{fig:soli_ampctr} the scalar amplitude at the
individual BS centres and the BS trajectories as functions of time
for the {\tt soli} and {\tt +soli} configurations starting with
initial separation $d=22.3\,M$.
Let us first consider the {\tt soli} configuration using plain
superposition displayed by the solid (black) curves. In the upper
panel of Fig.~\ref{fig:soli_ampctr}, we clearly see that the scalar
amplitude steadily increases, reaching a maximum around $t\approx
30\,M$ and then rapidly drops to a near-zero level.  Our interpretation
of this behaviour as a collapse to a BH is confirmed by the horizon
finder which reports an apparent horizon of irreducible mass $m_{\rm
irr}=0.5\,M$ just before the scalar field amplitude collapses; the
time of the first identification of an apparent horizon is marked
by the vertical dotted black line at $t\approx 30\,M$.  For reference
we plot in the bottom panel the trajectory of the BS centres along
their collision (here the $x$) axis. In agreement with the horizon
mass $m_{\rm irr}=0.5\,M$, the trajectory clearly indicates that
around $t\approx 30\,M$, the BSs are still far away from merging
into a single BH; in units of the ADM mass, the individual BS radius
is $r_{99}=2.78\,M$. We interpret this early BH collapse as a
spurious feature due to the use of plain superposition in the initial
data construction.  This behaviour is also seen in the case of the
real scalar field oscillatons in \cite{Helfer:2018vtq}.

We have tested this hypothesis with the evolution of the adjusted
initial data.  These exhibit a drastically different behaviour in
the collision {\tt +soli} displayed by the dashed (red) curves in
Fig.~\ref{fig:soli_ampctr}.  Throughout most of the infall, the
central scalar amplitude is constant, it increases mildly when the
BS trajectories meet near $x=0$, and then rapidly drops to zero.
Just as the maximum amplitude is reached, the horizon finder first
computes an apparent horizon, now with $m_{\rm irr}=0.99\,M$, as
expected for a BH resulting from the merger; see the vertical red
line in the figure.

As a final test of our interpretation, we compare the behaviour of
the binary constituents with that of single BSs boosted with the
same velocity $v=0.1$. As expected, the scalar field amplitude at
the centre of such a single BS remains constant within high precision,
about $\mathcal{O}(10^{-5})$, on the timescale of our collisions.
We have then repeated the single BS evolution by poisoning the
initial data with the very same term (\ref{eq:metricpert}) that is
also added near a single BS's centre by the plain-superposition
procedure.  The resulting scalar amplitude at the centre of this
poisoned BS is shown as the dash-dotted (blue) curve in
Fig.~\ref{fig:soli_ampctr} and nearly overlaps with the corresponding
curve of the {\tt soli} binary. Furthermore, the poisoned single
BS collapses into a BH after nearly the same amount of time as
indicated \setcounter{footnote}{4} by the vertical blue dotted curve
in the figure\footnote{ Recall that this BS model is stable but
fairly close to the stability threshold in Fig.~\ref{fig:statBS}
and therefore does not require a large perturbation to be toppled
over the edge.}.  Clearly this behaviour of the single boosted BS
is unphysical, and strongly indicates that the plain superposition
of initial data introduces the same unphysical behaviour to our
{\tt soli} binary constituents.  We have repeated this analysis for
our entire sequence of {\tt soli} binaries with very similar results:
the individual BSs always collapse to distinct BHs about $\Delta
t\approx 50\,M$ before the binary merger.

Finally, the trajectories in the bottom panel of Fig.~\ref{fig:soli_ampctr}
indicate that the BS merger occurs a bit later for the {\tt +soli}
case than its plain-superposition counterpart {\tt soli}. This is
indeed a systematic effect we see for all initial separations $d$
and which agrees with the different arrival times of the peak GW
signals that we have already noticed in Fig.~\ref{fig:soli_psi20}.
We do not have a rigorous explanation of this effect, but note that
the two trajectories in Fig.~\ref{fig:soli_ampctr} start diverging
right at the time of spurious BH formation in the {\tt soli} binary.
Perhaps some of the binding energy in BS collisions is converted
into deformation energy rather than simply kinetic energy of the
stars' centres of mass, slowing down the infall compared to the BH
case\footnote{We note that the relativistic Love numbers (which
measure the tidal deformability) of non-rotating BHs are zero
\cite{Binnington:2009bb}.}.  Another explanation may consider the
generally repulsive character of the scalar field which endows it
with support against gravitational collapse. When the infalling BSs
collapse to BHs, the scalar field essentially disappears as a
potentially repulsive ingredient and the ensuing collision is sped
up. Whatever ultimately generates this effect, the key observation
of our study is that even rather mild imperfections in the initial
data can drastically affect the physical outcome of the time
evolution.

%=============================================================================
\section{Conclusions}
\label{sec:conclusions}
We have simulated head-on collisions of equal-mass, non-spinning
boson stars and the GW radiation generated in the process.  The
main focus of our study is the construction of BS binary initial
data and the ensuing impact of systematic errors on the physical
results of the simulations. In particular, we have contrasted the
relatively common method of plain superposition according to
Eq.~(\ref{eq:superpossimple}) with the adjusted procedure
(\ref{eq:superposplus}) first identified in Ref.~\cite{Helfer:2018vtq}
for oscillatons.

Our results demonstrate that the adjustment (\ref{eq:superposplus})
in the construction of initial data leads to major improvements in
the initial constraint violations and the time evolutions of binary
BS collisions. In contrast, we find that the use of plain superposition
for BS binary initial data may not only result in quantitatively
wrong physical diagnostics but can even result in completely spurious
physical behaviour such as premature gravitational collapse. In
spite of the great simplicity of the adjustment (\ref{eq:superposplus})
and its success in overcoming the most severe errors in the ensuing
evolution, it is not free of shortcomings.  (i) In its present form,
the adjustment only works for a restricted class of binaries, namely
equal-mass systems with no spin and velocity vectors satisfying
$v_{\rm A}^i v_{\rm A}^j=v_{\rm B}^i v_{\rm B}^j$. (ii) Even with
the adjustment, the initial data contain some residual constraint
violations; it should therefore primarily be regarded as an improved
initial guess for a constraint solving procedure rather than the
``real deal'' in its own right.  These shortcomings clearly point
towards the most urgent generalisations of our work, overcoming the
symmetry restrictions and adding a numerical constraint solver.

%=============================================================================
\section*{Acknowledgments}
We thank Andrew Tolley, Serguei Ossokine and Richard Brito for
fruitful discussions.
This work is supported by
STFC Consolidator Grant Nos.~ST/V005669/1 and ST/P000673/1,
NSF-XSEDE Grant No.~PHY-090003,
STFC Capital Grant Nos.~ST/P002307/1, ST/R002452/1,
STFC Operations Grant No.~ST/R00689X/1 (project ACTP 186),
PRACE Grant No.~2020225359,
and
DIRAC RAC13 Grant No.~ACTP238.
Computations were performed on
the San Diego Supercomputing Center's clusters Comet and Expanse,
the Texas Advanced Supercomputing Center's Stampede2,
the Cambridge Service for Data Driven Discovery (CSD3) system,
Durham COSMA7 system and the Juwels cluster at GCS@FZJ, Germany.
T.H.~is supported by NSF Grants No. PHY-1912550 and AST-2006538,
NASA ATP Grants No. 17-ATP17-0225 and 19-ATP19-0051, NSF-XSEDE Grant
No. PHY-090003, and NSF Grant PHY-20043.  This research project was
conducted using computational resources at the Maryland Advanced
Research Computing Center (MARCC).  The authors acknowledge the
Texas Advanced Computing Center (TACC) at The University of Texas
at Austin for providing HPC resources that have contributed to the
research results reported within this paper. URL:
http://www.tacc.utexas.edu \cite{10.1145/3311790.3396656}

%=============================================================================
\section*{References}
%
%\bibliography{newuli2}
\bibliographystyle{unsrt}

\appendix

\section{Analytic treatment of the Hamiltonian constraint}
\label{sec:hamanalytic}
For the case of two non-boosted BSs, we can analytically compute
the Hamiltonian constraint violation at the stars' centres. Let us
consider for this purpose the metric ansatz (\ref{eq:ds2iso}). From
this line element, we directly extract the spatial metric
\begin{equation}
  \gamma_{ij}^{A} = \psi_{\rm A}^4 \delta_{ij}\,.
  \label{eqn:appendixansatzmetric}
\end{equation}
for a non-boosted BS at position $x_{\rm A}^i$.  This metric is
time-independent, so that for zero shift vector the extrinsic
curvature vanishes, $K_{ij}^{\rm A} = 0$.  For the second binary
member, we likewise obtain a metric $\gamma^{\rm B}_{ij}$ and
extrinsic curvature $K^{\rm B}_{ij}=0$, now centred at position
$x_{\rm B}^i$.

For sufficiently large initial separation $d=||x_{\rm B}^i -x_{\rm
A}^i||$, the exponential falloff of the scalar field implies
\begin{equation}
\begin{aligned}
    \phi_{\rm A}(x_{\rm B}) &= \phi_{\rm B}(x_{\rm A}) \approx 0\,, \\
     \Pi_{\rm A}(x_{\rm B}) &= \Pi_{\rm B}(x_{\rm A}) \approx 0\,.
\end{aligned}
\label{eq:apptmp1}
\end{equation}
The superposition of the two stars' scalar fields results in
\begin{align}
  &\varphi = \varphi_{\rm A} + \varphi_{\rm B}\,,
  \hspace{0.85cm}~~~~~~~~~~
  \Pi = \Pi_{\rm A} + \Pi_{\rm B}\,,
\end{align}
and, combined with Eqs.~(\ref{eq:apptmp1}),
\begin{equation}
\begin{aligned}
  \rho(x_{\rm A}) &= \rho_A(x_{\rm A})\,, \\
  \rho(x_{\rm B}) &= \rho_B(x_{\rm B})\,.
\end{aligned}\label{eqn:energydensityconditon}
\end{equation}
The single BS spacetimes are solutions to the Einstein equations;
by using Eq.~(\ref{eqn:appendixansatzmetric}), their individual
Hamiltonian constraints (\ref{eq:ham}) simplify to
\begin{equation}
    \mathcal{H}_{\rm A} = 8 \delta^{ij} \partial_i\partial_j
    \psi_{\rm A} +  16\pi \psi_{\rm A}^5\rho_{\rm A} = 0\,,
    \label{eq:hamA}
\end{equation}
and likewise for star B.

Next, we construct a binary spacetime by superposing the metric
which leads to
\begin{equation}
    \psi^4 = \psi_{\rm A}^4 + \psi_{\rm B}^4-c^4\,,
\end{equation}
where $c$ is a constant which we keep arbitrary for the moment.
For the Hamiltonian constraint of the superposed spacetime at the
centre of star A, we find
\begin{equation}
\begin{aligned}
    \mathcal{H}(x_{\rm A}^{i}) &=  8 \delta^{ij}
    \partial_i\partial_j{\psi_{\rm A}({x_{\rm A}^{i}})} + 8 \delta^{ij}
    \partial_i\partial_j{\psi_{\rm B}({x_{\rm A}^{i}})}\\ &
    +  16\pi \left[{\psi_{A}({x_{\rm A}^{i}})}^4
    + {\psi_{\rm B}({x_{\rm A}^{i}})}^4-c^4\right]^{5/4}\rho(x_{\rm A}^{i})\,.
\end{aligned}
\end{equation}
We can now choose the constant $c$ in accordance with the ``trick''
in Eq.~(\ref{eq:superposplus}), namely
\begin{equation}
   c = \psi_B(x_A^{i})\,,
\end{equation}
and the constraint simplifies to
\begin{equation}
\begin{aligned}
    \mathcal{H}(x_{\rm A}^{i}) &=
    8 \delta^{ij} \partial_i\partial_j{\psi_{\rm A}({x_{\rm A}^{i}})}
    + 8 \delta^{ij} \partial_i\partial_j{\psi_{\rm B}({x_{\rm A}^{i}})}\\ &
    +  16\pi {\psi_{\rm A}({x_{\rm A}^{i}})}^4 \rho(x_{\rm A}^{i})\,.
\end{aligned}
\end{equation}
By Eq.~(\ref{eq:hamA}), the derivative of the conformal factor
$\psi_{\rm A}$ cancels out the density $\rho_{\rm A}$, so that
\begin{equation}
\begin{aligned}
  \mathcal{H}(x_{\rm A}^{i}) &=
  8 \delta^{ij} \partial_i\partial_j{\psi_{\rm B}({x_{\rm A}^{i}})}\,.
\end{aligned}
\end{equation}
Using the analogue of Eq.~(\ref{eq:hamA}) for star B, we trade the
right-hand side for the energy density,
\begin{equation}
\begin{aligned}
  \mathcal{H}(x_{\rm A}^{i}) &=   - 16\pi {\psi_{\rm B}({x_{\rm A}^{i}})}^4
  \rho_{\rm B}(x_{\rm A}^{i})\,.
\end{aligned}
\end{equation}
For sufficiently large separation $d$ of the stars, however, this
vanishes by Eq.~(\ref{eq:apptmp1}) which is the result we wished
to compute. By symmetry, we likewise obtain $\mathcal{H}(x_{\rm
B}^i)=0$, which concludes our calculation.

\end{document}